\documentstyle[referee]{mn}
\newif\ifAMStwofonts

\input{psfig}
\input{epsf}

\ifoldfss
  \ifCUPmtlplainloaded \else
    \NewTextAlphabet{textbfit} {cmbxti10} {}
    \NewTextAlphabet{textbfss} {cmssbx10} {}
    \NewMathAlphabet{mathbfit} {cmbxti10} {} 
    \NewMathAlphabet{mathbfss} {cmssbx10} {} 
  \fi
  \ifAMStwofonts
    \ifCUPmtlplainloaded \else
      \NewSymbolFont{upmath} {eurm10}
      \NewSymbolFont{AMSa} {msam10}
      \NewMathSymbol{\upi}     {0}{upmath}{19}
      \NewMathSymbol{\umu}     {0}{upmath}{16}
      \NewMathSymbol{\upartial}{0}{upmath}{40}
      \NewMathSymbol{\leqslant}{3}{AMSa}{36}
      \NewMathSymbol{\geqslant}{3}{AMSa}{3E}

       \let\le=\leqslant
       \let\ge=\geqslant
    \fi
  \fi
\fi 

\ifnfssone
  \newmathalphabet{\mathit}
  \addtoversion{normal}{\mathit}{cmr}{m}{it}
  \addtoversion{bold}{\mathit}{cmr}{bx}{it}
  \newmathalphabet{\mathbfit} 
  \addtoversion{normal}{\mathbfit}{cmr}{bx}{it}
  \addtoversion{bold}{\mathbfit}{cmr}{bx}{it}
  \newmathalphabet{\mathbfss} 
  \addtoversion{normal}{\mathbfss}{cmss}{bx}{n}
  \addtoversion{bold}{\mathbfss}{cmss}{bx}{n}
  \ifAMStwofonts
    \ifCUPmtlplainloaded \else
      \UseAMStwoboldmath
      \makeatletter
      \new@mathgroup\upmath@group
      \define@mathgroup\mv@normal\upmath@group{eur}{m}{n}
      \define@mathgroup\mv@bold\upmath@group{eur}{b}{n}
      \edef\UPM{\hexnumber\upmath@group}
      \new@mathgroup\amsa@group
      \define@mathgroup\mv@normal\amsa@group{msa}{m}{n}
      \define@mathgroup\mv@bold\amsa@group{msa}{m}{n}
      \edef\AMSa{\hexnumber\amsa@group}
      \makeatother
      \mathchardef\upi="0\UPM19
      \mathchardef\umu="0\UPM16
      \mathchardef\upartial="0\UPM40
      \mathchardef\leqslant="3\AMSa36
      \mathchardef\geqslant="3\AMSa3E

       \let\le=\leqslant
       \let\ge=\geqslant
    \fi
  \fi
\fi 

\ifnfsstwo
  \DeclareMathAlphabet{\mathbfit}{OT1}{cmr}{bx}{it}
  \SetMathAlphabet\mathbfit{bold}{OT1}{cmr}{bx}{it}
  \DeclareMathAlphabet{\mathbfss}{OT1}{cmss}{bx}{n}
  \SetMathAlphabet\mathbfss{bold}{OT1}{cmss}{bx}{n}
  \ifAMStwofonts
    \ifCUPmtlplainloaded \else
      \DeclareSymbolFont{UPM}{U}{eur}{m}{n}
      \SetSymbolFont{UPM}{bold}{U}{eur}{b}{n}
      \DeclareSymbolFont{AMSa}{U}{msa}{m}{n}
      \DeclareMathSymbol{\upi}{0}{UPM}{"19}
      \DeclareMathSymbol{\umu}{0}{UPM}{"16}
      \DeclareMathSymbol{\upartial}{0}{UPM}{"40}
      \DeclareMathSymbol{\leqslant}{3}{AMSa}{"36}
      \DeclareMathSymbol{\geqslant}{3}{AMSa}{"3E}

       \let\le=\leqslant
       \let\ge=\geqslant
    \fi
  \fi
\fi 

\ifCUPmtlplainloaded \else
  \ifAMStwofonts \else 
    \def\upi{\pi}
    \def\umu{\mu}
    \def\upartial{\partial}
  \fi
\fi

\title{The evolution of a supermassive binary caused by an
accretion disc }
\author[P.B.Ivanov, J.C.B. Papaloizou 
and A. G. Polnarev]
       {P.B.Ivanov$^{1,2,3}$\thanks{E-mail:pavel@tac.dk}, 
J.C.B. Papaloizou$^{1}$\thanks{J.C.B.Papaloizou@qmw.ac.uk} 
and A. G. Polnarev$^{1,3}$\thanks{A.G.Polnarev@qmw.ac.uk}
 \\$^{1}$Queen Mary $\&$ Westfield College, University of
London, London E1 4NS, United Kingdom
\\$^{2}$ Teoretical Astrophysics Center,
Juliane Maries Vej 30, DK-2100 Copenhagen ${\o}$, Denmark\\
$^{3}$ Astro Space Center of P.N. Lebedev Physical Institute of the Russian
Academy of Sciences,\\
Profsoyuznaya st., 84/32, Moscow 117810, Russia}
        
\pagerange{\pageref{firstpage}--\pageref{lastpage}}
\pubyear{}

\begin{document}

\maketitle

\label{firstpage}
\begin{abstract}
 The interaction between a massive binary and a  non-self-gravitating 
circumbinary accretion disc
 is considered. The shape of the stationary 
twisted disc  produced by the binary
is calculated.
It is shown that  the inner part of the disc must lie in the
binary orbital plane for any value of the ivi@tfa.fy.chalmers.se viscosity.

 When the inner 
 disc midplane is
aligned with the binary orbital plane  on the scales of interest and
it rotates  in the same sense as the binary, the modification to the disc
structure and  the rate of decay of the  binary orbit, assumed circular,  due to
tidal exchange of angular momentum
 with the disc, are calculated. It is shown
that the modified disc structure is well described by  a self-similar
solution of  the non-linear  diffusion equation   governing the  evolution of the disc surface
density. The calculated time scale for decay of the binary 
orbit is always smaller than ``accretion'' time scale 
$t_{acc}=m/\dot M$ ($m$ is the mass of secondary component, 
and $\dot M$ is the disc accretion 
rate), and is determined by ratio of secondary mass $m,$ assumed to be much
smaller than the primary mass,   
the disc mass inside the initial 
binary orbit, and  the form of viscosity in the
disc.   
\end{abstract}


\section{Introduction}
In the present paper we discuss effects of interaction 
between a massive binary and circumbinary accretion disc.  We
consider in detail the evolution of the binary orbital
parameters and the disc surface density.

 A massive binary  surrounded by a gas disc may be found
in different astrophysical situations. For example, such a system
may contain two pre-main-sequence stars surrounded by a protostellar 
disc. The tidal influence of this disc on the   binary ormit may cause 
orbital migration, and in this way explain the 
observed small separations of some  pre-main-sequence  
binaries.  Similar processes may operate in protoplanetary
discs   producing short period planets. The influence of the binary on the 
disc may be to significantly modify the disc structure by opening  up a gap.
 The effect of  such a gap may may be  to reduce the total  luminosity
of the disc, and change its spectrum, and therefore it has
observational consequences (see, e.g.
Jensen et al 1996; Jensen $\&$ Mathieu 1997 for
recent observations and an overview of theoretical
models).

Probably  an even more important example is a
supermassive  binary black hole which is expected to be found in
the centers of some galaxies. Such a system may be formed as a
result of a galaxy merging process, and  the evolution may end  with 
 the coalescence of the black holes. The  burst of gravitational waves 
generated during such  a coalescence   should be easily detectable   by  the future
generation of gravitational wave antennas.   Just as in the case of 
binary stars, the interaction of a binary black hole with an
accretion disc has prominent evolutionary aspects. The disc is able
to collect and radiate away  orbital binding energy, and
thus takes part in  the  process of decay  of the black hole
separation distance. 

 In fact the  disc orbit interaction may 
determine the evolution 
of the binary separation distance  over  a definite range of scales.
To see this let us  briefly review the possible   
 causes of  evolution of 
the separation distance (Begelman, Blandford $\&$ Rees 1980,
hereafter BBR). After the  merger of the galaxies,  the holes 
sink toward the center of  of the stellar distribution (central stellar
cluster) due to dynamical
friction.   When they are 
  within a small  distance from the center $\sim 1-10pc$ they 
form a gravitationally bound binary. After that the binary
separation distance $r_{b}$ shrinks  as a result of dynamical
friction to another smaller 
characteristic scale $r_{crit}\sim 10^{-2}-1pc$ where
the binary becomes ``hard'' with respect to
stellar population, i.e. the velocity obtained by a star during 
some typical encounter with the binary becomes 
larger than escape velocity of stars from the stellar cluster. 
Thus, a ``loss cone'' is
formed in  the velocity distribution of  the stars in the cluster, 
and the binary can
effectively interact only with stars which penetrate into ``loss cone''
 at a rate determined by star-star relaxation processes.

  For separations
$r_{b}\sim 10^{-2}-1pc$
decay of the binary orbit due to emission of 
gravitational waves appears to be insignificant,
and the decay
time scale is mainly determined by  the two-body relaxation time
of  the  central stellar cluster,
  which may be much larger than  the cosmological time (BBR, see also
Polnarev $\&$ Rees
1994; Quinlan 1996;
Quinlan $\&$ Hernquist 1997 and references therein
for  a more recent discussion).

Note that the successive encounters of the stars 
with the binary lead also to  a ``random walk''
of the binary center of mass around the center of stellar distribution
(Bahcall $\&$ Wolf 1976; Young 1977), and this effect may ``smear
out'' the loss cone (Young 1977; Quinlan $\&$ Hernquist 1997).
Although the problem 
of speeding up the  orbital evolution of the binary might be solved in such a way, 
it would be very plausible to  suppose there is another mechanism 
 that can produce  binary orbital energy loss at
``intermediate'' scales $\sim 10^{-2}-1pc$. In this connection  we
 note that the size of the thin accretion disc around the  more
massive black hole (primary) may be as large as $\sim 10^{-2}-10^{-1}pc$,
and therefore the interaction between the binary and the disc
 may ``switch on'' at the most critical ``intermediate'' scales
$\sim 10^{-2}-10^{-1}pc$.      

The black hole binary-disc 
interaction can also have significant observational
consequences in its own right. As we shall see in the next Section,
during the initial stage of the binary-disc interaction the
orbit of  the smaller (secondary) black hole can be inclined with
respect to the circumprimary disc, and in this case the secondary
hits the disc twice during its orbital period. At the time of
secondary-disc collision  outflows of hot gas from the disc
are formed, and the luminosity of the  outflowing gas can be of
order of secondary Eddington luminosity (Ivanov et al 1998, 
hereafter IIN). Therefore the binary black hole may manifest itself as 
a periodic source of non-stationary radiation coming from
the centers of galaxies (see e.g. INN; Lehto $\&$ Valtonen
1996 and references therein). 
As we see below, during the final stage of the binary-disc 
interaction the binary
orbital plane and the disc plane are aligned, and  as for 
protostellar discs,  the disc structure is modified by the 
gap formation
process. The specific spectral features  expected from
 such a modified disc can also be used to  test
the hypothesis of supermassive binary black holes.

In this Paper we consider the evolutionary aspect of the interaction
between a massive binary system (presumably, a binary black hole) and
an accretion disc. We assume that our system consists 
of two companions with nonequivalent masses: a massive primary with
mass $M$, and  
a secondary with mass $m \ll M$; it also contains
an accretion disc around the primary. We also assume that in course
of previous evolution the binary separation distance $r_{b}$ becomes
smaller than the disc size $r_{d}$, so the disc mass inside the binary
orbit $M_{d0}(r_{b}) \le  M_{d0}(r_{d})$, where $M_{d0}(r_{d})$ is
total mass of the disc 
(sometimes we shall call the distance $r_{b}$  the orbital distance
of the secondary). We pay special attention to the case of a  rather massive 
secondary: $M_{d0}(r_{b})< m \le M_{d0}(r_{d})$. Such a  massive secondary
significantly disturbs the inner part of the disc, and causes the
evolution of the disc into a  disturbed quasistationary state during a  
relatively short time-scale. The subsequent evolution of the binary
parameters takes a  much longer time, and can be calculated for a given
perturbed state of the disc.

In the beginning  stages of the binary-disc interaction,
the binary orbital plane 
is not necessary aligned with the disc plane, 
and therefore the gravitational field of
the binary leads to  disc distortions of two types:  ``horizontal''
twisted distortions of the disc shape due to action 
of the non-spherical part of the binary gravitational potential, 
and ``radial'' distortions of the disc structure 
due to the  tidal torque  from the secondary
acting on the disc gas particles which are orbiting 
close to the binary orbit. The ``horizontal'' distortions cause the
formation of a  quasi-stationary twisted disc where the inner part  
of the disc ( with radius smaller than some alignment 
radius $r_{al}$) lies in the binary orbital plane, and the orientation
of the outer part of the disc is determined by the disc formation
processes (see also Section 2,3). We calculate the
shape of this twisted disc, and discuss the dependence of
alignment scale on viscosity and mass ratio $q=m/M$ (see also 
Kumar 1987,1990). The tidal torque of the secondary
 inhibits the disc gas  from drifting inside the secondary orbit
(Lin $\&$ Papaloizou 1979 ; Goldreich $\&$ Tremaine 1980), and   so it   
 accumulates in some region outside the secondary orbit,
and the radial structure of the disc 
in that region is modified (e.g. Syer $\&$ Clarke 1995).
We calculate analytically 
the disc structure
in that region in two ways: a)
 by using the laws
of conservation of mass and angular momentum and b) by finding
 the appropriate self-similar solution of  the diffusion equation
which governs  the viscous evolution of  the surface density in the
disc, and comparing  the results  with  the
results of simple numerical calculations. 
We also calculate the evolution of the secondary orbital
 separation $r_{b}(t)$ due to the torque acting  between the modified
part of the disc and the secondary. 

Our paper is organized as follows. In the next Section we estimate
the characteristic time and spatial scales relevant 
for interaction between the binary and the disc, 
and qualitatively describe the possible evolution of such a system. 
All parameters are assumed to be typical for  a supermassive binary
black hole and galactic accretion disc. In Section 3 we calculate
the shape of  the stationary twisted disc around an inclined binary.
In Section 4 we calculate the ``radial''
evolution of the binary and the disc. We summarize our results
in Section 5. 
Although we mainly discuss the case of  a supermassive binary
black hole, our results can be applied to other possible
situations such as stellar binaries or protoplanetary discs, with minor
changes. The reader who is mainly interested in twisted discs around
binary systems, can  go  directly
to  Section 3, and the reader who is mainly
interested in  the  general aspects of the ``radial'' evolution
of the binary and the disc, can jump to  Section 4.

We consider hereafter only circular binaries, and set eccentricity
$e$ to zero in our calculations.    

\section{Main parameters and basic estimates}

In the beginning of this Section let us shortly review 
the basic properties
of steady-state accretion discs.
The detailed disc model is 
unimportant for us, and we employ 
simple $\alpha$-disc models (Shakura $\&$ Sunyaev 1973) 
which are parameterized by the well known viscosity parameter $\alpha$
and accretion rate $\dot M$.  For a typical AGN disc
the values of viscosity and  dimensionless accretion
rate are assumed to be small: $\alpha \sim 10^{-2}$,
$\dot M /\dot M_{E} \sim 10^{-2}$, where $\dot M_{E}$
is the Eddington accretion rate 
$$\dot M_{E}={4\pi GM\over c\kappa_{T}}\approx 
2\times 10^{-1}M_8(M_{\odot}/yr), \eqno 1$$
here $\kappa_{T}$ is Thompson opacity coefficient, 
and $M_{8}=M/10^8M_{\odot}$, and
all other symbols have their usual meaning. 
We use later renormalized
values of  viscosity and accretion rate:
$\alpha_{*}=\alpha/10^{-2}$, and 
$\dot M_{*}=10^{2}\dot M/\dot M_{E}$.  

All interesting disc quantities can easily be expressed
in terms of disc  semi-thickness $h$, and steady state
surface density $\Sigma_0$. 
In the gas pressure dominated case, 
the   semi-thickness is
approximately proportional to the distance $r$, 
and to describe the vertical extent of the disc
it is convenient to introduce 
the disc opening angle: 
$$\delta 
={h\over r}
\approx 10^{-3}
\alpha_{*}^{-1/10}
M_{8}^{-1/10}{\dot M_{*}}^{1/5}
{(r_{3})}^{1/20}, 
\eqno 2$$
where the distance $r_{3}={r\over 10^{3}r_{g}}$
is assumed to be
of order of the secondary orbital distance $r_{b}$, and 
gravitational radius $r_{g}=2GM/c^{2}$. 
\footnote{In this Section we assume that opacity law in the disc
is determined by Thompson opacity. 
The expressions for other reasonable
opacity laws give similar results (see eqs. (44-47)  for a general
case).} 
The steady state disc surface density is given by expression
$$\Sigma_0
\approx
10^{5}\alpha_{*}^{-4/5}M_{8}^{1/5}{\dot M_{*}}^{3/5}r_{3}^{-3/5}
g/cm^{2}. \eqno 3$$ 
With help of the expression (3) we can easily estimate the unperturbed
disc mass $M_{d0}$ inside the secondary orbit
$$M_{d0}(r_{b})\approx 
{10\over 7}\pi 
\Sigma_0 (r_{b})r_{b}^{2}\sim 2\cdot 10^{5}\alpha_{*}^{-4/5}
M_{8}^{11/5}{\dot M_{*}}^{3/5}r_{3}^{7/5}M_{\odot}. \eqno 4$$
The disc gas particles are orbiting around the central source with
approximately Keplerian angular velocity
$$\Omega_K=\sqrt{{GM\over r^3}}\approx 
0.7\cdot M_{8}^{-1}r_{3}^{-3/2}{(yr)}^{-1}, \eqno 5$$
and slowly drifting toward the center due to the action of viscous 
forces. The characteristic ``viscous''
time scale follows from  eqs. (1,4):
$$t_{\nu 0}={M_d\over \dot M}\sim  10^{8}\alpha_*^{-4/5}
M_8^{6/5}{\dot M_{*}}^{-2/5}r_{3}^{7/5}yr \eqno 6$$
Sometimes we also use the simple estimate for the drift
time scale:
$$t_{\nu 0}\sim \alpha^{-1}\delta^{-2}\Omega_K^{-1}. \eqno 7$$

The size of a thin 'standard' disc $r_{d}$ may be determined by
the specific angular momentum of the gas which  enters the disc,
or by processes of star formation in the outermost regions 
of the disc, or any  processes which can 'turn off' 
accretion. For the ``standard'' accretion disc with
value of viscosity parameter $\alpha \sim 10^{-2}$, the
Toomre stability parameter $Q$ can be rather small:
$$Q\approx {\Omega_{K} c_{s}\over \pi G\Sigma}\approx
{\delta M\over M_{d0}}\approx 0.5\alpha_{*}^{7/10}
M_{8}^{-13/10}\dot M_{*}^{-2/5}r_{3}^{-27/20}$$
In the disc region where $Q <1$ the disc is  gravitationally unstable
 and locally gas may
 collapse to  form low-mass stars (e. g. Spitzer
$\&$ Saslaw 1966, Illarionov $\&$
Romanova 1988). 
We do not consider such a disc here  as we
 adopt disc parameters such  that $Q \ge 1.$
As  mentioned above, 
we consider the case when the interaction between the secondary   
and the disc occurs on scales smaller
than the disc size $r_{d}$. That condition can be described
in terms of characteristic 
radial scale $r_{m}$ -the scale where the disc mass is  
equal to the mass of the secondary:
$$r_{m}\approx 3\cdot 10^{3} \alpha_{*}^{4/7}m_{6}^{5/7}M_{8}^{-11/7}
{\dot M_{*}}^{-3/7}r_{g}$$
where $m_{6}=m/10^{6}M_{\odot}$.
The secondary interacts only with gas contained inside the scale
$r_{m}$, and the disc region outside this radius is in 
the unperturbed steady state 
\footnote{Actually, in case of a very massive secondary,
the disc structure is close to the steady state solution at scales
smaller than $r_m$, see eqs. (63,64) below.}. 
The gas contained inside 
the radius $r_{m}$ is accreted during the time 
$t_{acc}={m\over \dot M}$:
$$t_{acc}=t_{\nu 0}(r_{m})=5\cdot 10^{8}{m_{6}\over M_{8}}{\dot M_{*}}^{-1}yr,
\eqno 8$$
and this time is larger than any characteristic time scale
relevant to our problem.
In the case of a supermassive binary black
hole, the 'accretion' time (8) must be smaller than cosmological time in 
order to provide the source of gravitational radiation (BBR).

Now let us describe qualitatively the scenario 
for  binary evolution due to the  interaction with the  accretion disc, 
and estimate the relevant characteristic
spatial and time scales. As we will see the interaction of the secondary 
with the disc gas leads to alignment of the secondary orbital plane
and the plane of the  inner disc. After the disc alignment has 
occurred, the orbital
separation starts to shrink in the radial direction. 
Therefore the evolution of the
binary and  disc can be roughly divided into two successive stages
-the alignment stage and the stage of 'radial' evolution, and we
consider these stages in turn.

First we assume that the secondary orbital plane
is inclined with respect to the disc plane. The effects
of the interaction between the inclined 
secondary and the disc can be  considered in two regimes. 
The first is when  highly
non-stationary distortions of the disc   result from  direct
collisions between the secondary and  disc (s.-d. 
collisions). The second is when  the long-ranged, averaged over many orbital
periods, interaction
between the binary and the disc  produces a twisted warped disc. 

As we will see the evolution of the secondary
orbit due to s.-d. collisions is important for relatively
low-mass secondaries, and long-ranged interaction
is important in the opposite case.  

Let us estimate effects of the direct
collisions between the secondary and the disc. The inclined
secondary  hits the
disc twice during  an orbital period $t_{orb}$,
which is approximately equal to the Keplerian  period:
$$t_{orb} \sim 2\pi \Omega_K^{-1}. \eqno 9$$ 
During 
a collision, the secondary significantly perturbs
the disc gas at  distances from it  comparable with
the ``accretion'' radius:
$$r_{a}={2Gm\over|\vec v_{sec}-\vec v_{d}|^{2}}
\approx 3\times 10^{14}m_{6}r_{3}cm, \eqno 10$$
where $\vec v_{sec}$ is the velocity of the secondary which
 intersects
the disc at the ``collision'' radius $r_c \sim r_b,$ 
$\vec v_{d}$ is the disc velocity at the point of  
collision, and
$|\vec v_{sec}|\sim |\vec v_{d}|\sim r_c\Omega (r_c)$.  
The perturbed
gas is heated up   by the  shock  induced in the disc. 
An amount of gas of  mass $ \sim \pi \Sigma r_a^{2}$ 
is accreted
by the secondary, another  amount of approximately the same
mass attains  velocities greater than  the escape velocity
in the binary potential and leaves the system (IIN).
As result of the  collision, the linear momentum of the secondary is
changed,   with the momentum change $\delta \vec p_{sec}$
 being proportional to the total mass of perturbed region
$ > 2\pi \Sigma (r_{b}) r_{a}^{2}$,  and it is directed opposite
to the relative 
velocity of the secondary $\vec v_{sec} -\vec v_{d}.$
Therefore the change of velocity  of the secondary
$\delta \vec v_{sec}$
can be written as:
$$\delta \vec v_{sec}
\approx -{2\pi \Sigma r_{a}^{2}A\over m}(\vec v_{sec} -\vec v_{d}),
\eqno 11$$
here the correction factor $A >1$ is mainly determined by the
disc distortions 
at scales larger than $r_{a}$
\footnote{We do not take into account the resonant excitation
of waves in the disc due to the secondary passing through the disc.
Estimates of this effect have been made e. g. by Artymowicz  1994.}.
The secondary orbital parameters  change  on account of the change
of velocity, and this leads to the dragging of the secondary
orbit into the disc plane. To see this, consider the evolution of
the inclination angle $\beta$: the angle between the specific
angular momentum of the secondary $\vec l_{sec}$, 
and the specific angular 
momentum  vector of the disc gas $\vec l_{d}= \vec r_c \times \vec v_d$
(see Fig. 1 below). 
From  eq. (11) it is easy to  
obtain the equation for the change of this angle during
the collision
\footnote{In the analogous case of star-disc 
collision,  similar calculations
and references to earlier work can be found in  a paper
by  Vokrouhlicky $\&$ Karas 1998.}:
$$\delta \beta=-{2\pi \Sigma r_{a}^{2}A\over m}
{|\vec l_{d}|\over |\vec l_{sec}|}\sin (\beta ). \eqno 12$$
One can easily see from this expression that there is
only one stable value of $\beta$: $\beta =0$.   
Thus,  collisions between the secondary and the disc tend
to align the angular momentum  vectors of the disc
and the secondary. When these collisions determine the 
evolution of the secondary orbital parameters, the secondary
is    progressively dragged into the disc plane   until it   is in a coplanar orbit.
The associated timescale $t_{dragg}$ 
 can be easily estimated with help of
eq. (11) to be
$$t_{drag}\approx {m\over 2\pi \Sigma r_{a}^{2}A}t_{orb}. \eqno 13$$

As we mentioned above the b.h-disc collisions result in
an outflow of  disc gas  at a rate
$$\dot M_{out} > \Sigma_{0}r_{a}^{2}\Omega_K, \eqno 14$$
Obviously, this outflow must be compensated
by a radial inflow of mass through the disc, and therefore the
surface density in  eqs. (11-13) is close to  the steady state
value (3) only if the 
condition $\dot M_{out} < \dot M$ is satisfied.
This condition gives:
$$q < \sqrt{{3\pi\over 4}}
\alpha^{1/2}\delta \sim 1.5\cdot 10^{-4}B\alpha_{*}^{1/2}, \eqno 15$$
where black holes mass ratio
$$q={m\over M},$$
the parameter $B=\delta/10^{-3}\approx 1$, and we use the standard
equation of stationary disc accretion: $\dot M=3\pi\nu\Sigma_{0}$
and kinematic viscosity $\nu =\alpha \delta^{2}r^{2}\Omega_{K}$. 
In the opposite case $q > \alpha_{*}^{1/2}\delta $, the secondary 
 depletes  gas  from the region of the disc near its
orbit, so the surface density in this region should be much less
than the unperturbed surface density $\Sigma_{0}$. Now  an upper limit for the 
surface density $\Sigma (r_{c})$ may be estimated by 
 equating the rate of outflow of  disc gas  
$\sim \Sigma (r_{c}) r_{a}^{2}\Omega_K $, and the disc accretion rate
$\dot M$, and that gives
$$ \Sigma (r_{p}) \sim {\alpha \delta^{2}\over q^{2}}\Sigma_{0}
\ll \Sigma_{0} \eqno 16$$   
In the limit ${\alpha \delta^{2}\over q^{2}}\ll 1$
the drag timescale is of order of 'accretion' time
scale $t_{acc}$.

In the case of a sufficiently massive binary, the disc
midplane is expected to be 
aligned with the binary orbital plane due to the influence of
quadrupole component of the binary gravitational field.
Schematically this process can be divided again into two 
successive stages. 
The planar disc  first evolves into a quasi-stationary
twisted configuration (the twisted disc) 
during some characteristic alignment time
$t_{al}$. In the twisted disc, the inner part of the disc 
is aligned with the binary  orbital plane, and the outer part does not
change its orientation.
The radius out to which the disc is aligned
($r_{al}$), as well as the alignment 
time $t_{al}$, are estimated in 
the next Section (see also Kumar 1987, 1990).  
Both quantities strongly 
depend on the value of viscosity parameter 
$\alpha$. 
For sufficiently large values of $\alpha > \delta$, we obtain:
$$r_{al1}\sim {(\alpha q)}^{1/2}\delta^{-1}r_{b}, \eqno 17$$
and diffusion-like decay to a quasi-stationary twisted disc
proceeds during the time
$$t_{al}\sim \alpha^{2}t_{\nu 0} \sim \alpha \delta^{-2}\Omega_K^{-1}.
\eqno 18$$
For small values of $\alpha$, the alignment scale does not
depend on viscosity (Ivanov $\&$ Illarionov 1997, hereafter II)
$$r_{al2}\sim q^{1/2}\delta^{-1/2}r_{b}, \eqno 19$$ 
and the evolution to the stationary configuration has wave-like
character (Papaloizou $\&$ Lin 1995, hereafter PL). 
Now the alignment time $t_{al}$ is 
approximately equal to the 'sound' time $t_{s}$: the time taken
by sound wave to cross the radius $r$:
$$t_{s}\sim \delta^{-1}\Omega_K^{-1}(r_{al}). \eqno 20$$  
Obviously, the alignment scale must be greater than the binary
orbital distance: $max(r_{al1}, r_{al2}) > r_{b}$. That condition
gives:
$$q > {\delta^{2}\over \alpha}\quad when \quad \alpha > \delta;
\quad q > \delta \quad when \quad \alpha < \delta. \eqno 21$$
When the inequalities (21) are broken, the alignment effect
is absent. 

After the formation of a quasistationary twisted disc,
the orientation of the binary orbital plane itself  slowly 
changes  with time due to the influence of the gas
accreting  through the twisted disc. When  the gas
is accreting  through the twisted disc, the orientation of its
angular momentum is changing, and therefore 
the orientation of the binary orbital plane must also be change
with time due to the law of conservation of angular momentum
(e. g. Rees 1978).
This fact allows us to estimate 
the change with time of the binary inclination angle $\beta_{out}$ 
as:
$$\dot \beta_{out} \sim -{\beta_{out} \over t_{*}}, \quad
\beta_{out} (t)\propto e^{-{t\over t_{*}}}, \eqno 22$$
where index $out$ means that the inclination is determined with
respect to the orientation of the disc  at radii far from the binary. 
The characteristic time-scale of this effect $t_{*}$ 
is determined by
the ratio of specific angular momentum of the binary to
the specific angular momentum of the disc  at radius
$r\sim r_{al}$, 
and by the 'accretion' time $t_{acc}$:
$$t_{*}\sim {r_{b}v_{d}(r_{b})\over r_{al}v_{d}(r_{al})}t_{acc}. 
\eqno 23$$
Here the alignment scale is given by eq. (17) and $\alpha$ is
assumed to be rather large ($\alpha \sim 1$) (the case of small 
$\alpha \ll 1$ is more complicated and must be considered separately,
see e.g. Scheuer $\&$ Feiler 1996).
When the secondary  counter rotates  with
respect to the disc gas ( retrograde rotation),
we have ${v_{b}\over v_{d}} < 0$, and the time $t_{ev}$ is formally
negative.
This means that 
the inclination of the 
binary orbital plane with respect to the disc plane
grows with time, 
and the binary may turn over and become prograde
during the time $\sim t_{ev}$. Note, however
that  a highly inclined twisted disc may be unstable with respect to
instabilities of a  different kind, and  so  the  evolution
of  a retrograde inclined binary is unclear. 
We will assume below
that the binary is rotating prograde with respect to the disc gas.

From  eq. (23) and  eqs. (18, 20) it follows that 
$t_{ev} \gg t_{al}$ when $m \gg M_{d}$. This inequality as well
as the inequalities (21) tell us that 
the disc alignment effects are
important only for sufficiently massive secondaries (typically
$q > 10^{-3}$). In the opposite case the secondary is quickly
dragged into the disc midplane by the competing 
effects of  secondary-disc collisions.

Now let us assume that the binary
and the disc are aligned at scales of interest,
and consider the next 'radial' stage of the  
evolution of the binary and the disc.
At first let us look at the case of the small  
mass of the secondary $m < M_{d}$. 
As we mentioned above a low massive secondary must
lie in the disc plane corotating with the disc gas. Orbiting
in the disc the secondary transfers the angular momentum to
the neighboring gas particles which are orbiting upstream,
and gain the angular momentum from downstream gas particles
due to resonance interactions/ non-resonance scattering of these
particles (Lin $\&$ Papaloizou 1979
Goldreich $\&$ Treimain 1980). 
This can lead to the formation of gap around the secondary
which is free from the gas, provided the gap formation
criterion is satisfied (see, e.g. Lin $\&$ Papaloizou 1979, 1986)
$$q > {81\pi \nu \over 8r_{b}^{2}\Omega_K}, \eqno 24$$
where $r_{b}$ is the binary separation distance.
For $\alpha$-discs the criterion (24) can be rewritten as
$$q > {81\pi \over 8}\alpha \delta^{2}\sim 3\times 
10^{-7}\alpha_{*}.$$
One can see from this relation that the criterion is 
obviously satisfied for any reasonable mass of the secondary.
Lin and Papaloizou showed
that when the gap is opened 
the orbit of secondary is shrinking 
during viscous time scale $t_{\nu 0}$
\footnote{Note that 
the secondary can shrink its orbit even though
it is not massive enough to open
a gap (Ward 1997 and references therein)}. 

At the end of this section
let us briefly discuss the second possibility $m > M_d$
( this case is discussed in detail in Section 3).  Now the
radial drift of the secondary proceeds during the time much
larger than $t_{\nu 0}$. The secondary torque acting on the 
downstream gas particle prevents them to drift inside the 
secondary orbit, and the downstream gas is accumulated in some
region outside
the secondary orbit. The structure of the disc in that region
is significantly modified. The radius out to which the disc is 
modified grows with time (see Section 3). The accumulated  
mass of the gas near the secondary orbit increases the 
torque acting on the secondary, and pushes the secondary 
toward the primary. The secondary orbit is shrinking
during the characteristic evolution 
time-scale $t_{ev}$ ($t_{\nu 0} \ll
t_{ev} < t_{acc}$), which is evaluated in the Section 3.

\section{The twisted accretion disc around the inclined binary}

Now let us consider in more detail 
the evolution and the structure of the twisted disc
around the binary.
At first we need to choose a convenient
coordinate system. As a basic coordinate system 
we use the Cartesian 
coordinate system $(x, y, z)$, which has its origin (O) at the
center of mass of the binary. The plane (XOY) coincides with
the binary orbital plane and the axis (OZ) is directed 
perpendicular to that plane. As it follows from our previous
discussion the binary orbital plane is slowly evolving with
time, but 
when considering the dynamics
and 
\clearpage
the structure of the twisted
disc itself the effect of this evolution is unimportant. 
Since the twisted disc has 
some inclination with respect to the binary plane, and this 
inclination is different for different radii $r$, when
studying the local properties of the disc it is convenient 
to use another locally cylindrical (twisted) coordinate
system $(r,\psi, \xi)$, (Petterson 1977).
This system consists of  rings centered at the point (O),
which have different (with radius) inclination angles 
$\beta (r,t)$ with respect to
the binary orbital plane, and the lines of nodes of these rings
are rotated  through some angle $\gamma(r,t)$
with respect to the (OX)-axis, see Fig. 1.
\begin{figure}
\parbox{20cm}
\centerline{
\psfig{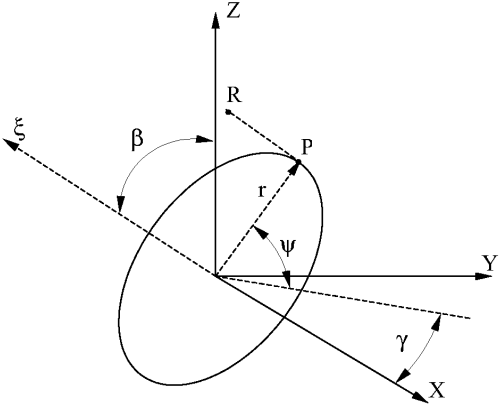}}
\caption{The $r$, $\psi$, $\xi$ coordinate system shown with respect
to the Cartesian coordinate system. $P$ is the projection of the point
$R$ on to the disc ring plane.}
\end{figure}
The position of the gas elements on the ring is characterized
by the angle $\psi$, and the coordinate $\xi$ characterizes the
vertical displacement 
of the gas element out of the ring plane (see Fig. 1).
\thispagestyle{plain}
Assuming that the twisted disc is evolving  on a  time scale much
larger than the orbital time scale, and the spatial scales of the
disc twist are much larger than the binary orbital distance, we
hereafter
use the time-averaged over many orbital periods
monopole and quadrupole 
components of the binary potential $\phi (r, \psi, \xi)$:
$$\phi(r, \psi, \xi)=-{GM\over R}-{Gmr_{b}^{2}\over 2r^{3}}
({3\over 2}{R^{2}\over r^{2}}(1-{2\xi \beta \sin \psi \over R}) 
-1), \eqno 25$$
where the spherical coordinate $R=\sqrt {r^{2}+\xi^{2}}$, and
we assume $\beta < 1$
\footnote{In the expression (25) the
assumption $m \ll M$ can be relaxed. 
The potential (25) describes 
the time-averaged gravitational field of the circular
binary with components having the comparable masses $m_{1}$,
$m_{2}$, provided $M=m_1 +m_2$, and $m={m_1 m_2\over m_1 +m_2}$.}.

When disc twist is small $(\beta \ll 1)$, trajectories
of the gas particles in the twisted disc are ellipses with 
small eccentricity $e$. The influence of the 
non-spherical part of the potential (25) on these trajectories
can be described as the following. Neglecting the
interaction between gas particles and then considering 
the free motion of these particles in the potential
(25), one can see that the  quadrupole
 contribution to the
potential causes  the  precession of the major axis of 
an elliptical  orbit with the frequency:  
$$\Omega_{ap}={3\over 4}q{({r_{b}\over r})}^2\Omega_{K},
\eqno 26$$
and  precession of the lines of nodes
with the frequency:
$$\Omega_{np}=-\Omega_{ap}. \eqno 27$$
The apsidal precession is prograde and the nodal precession
is retrograde with respect to the orbital motion.
As we will see,  the  signs of these precession frequencies 
determines the shape of the stationary twisted disc in the
low-viscosity limit.

The dynamics and the structure of a twisted disc  can be described
in terms of two complex variables: ${\bf W}(r,t)=\beta e^{i\gamma}$, and
${\bf A}(r, t)$. The variable ${\bf W}$ determines the inclination and
the rotation of the disc rings
\footnote{The variable ${\bf W}$ is connected to the variable $g$ used by
PL by ${\bf W}={(ig)}^{*}$.}, and the variable 
${\bf A}(r, t)$ determines
the deviation
of the gas trajectories from the circle of the radius $r$. 
Namely,
the eccentricity $e$ and the position of the  apsidal line of the
elliptical trajectory $\psi_0$ can be expressed in terms of ${\bf A}$:
$e=-2{\xi \over r}\Omega_{K}|{\bf A}|$, $\psi_0=\arccos {Re {\bf A}\over |{\bf A}|}$
(Demianski $\&$ Ivanov 1997).
In the low-viscosity limit $\alpha < 1$, the dynamical twist 
equations can be written as
$${\dot {\bf A}}=-{GM\over 4r^{2}}{\bf W}_{,r}+
(i\Omega_{ap}-\alpha \Omega_{K}) {\bf A}, \eqno 28$$
$${\dot {\bf W}}=-{\delta^{2}\over r}{\partial \over \partial r}
(r^{2}{\bf A})+i\Omega_{np}{\bf W}. \eqno 29$$

The derivation of these equations lies beyond the scope of our paper,
but can be found in  papers by Papaloizou and  Pringle 1983, Papaloizou and
Lin 1995, Ivanov and Illarionov 1997, Demianski and Ivanov 1997. 
Note that the equations
 (28-29) are applicable not only to the case of a binary system,
but also for  general perturbing forces which can cause  apsidal and
nodal precession of the disc gas trajectories in the linear approximation
over the angle $\beta$ (II, Demianski $\&$ Ivanov 1997).

We do not solve the time dependent form of these equations in our work, and
restrict ourselves to  qualitative analysis only
\footnote{For the discussion of numerical solutions of these equations see Kumar 1990,
PL.}.
For simplicity in our analysis of the time dependent case we also 
neglect the terms proportional to the precession frequencies 
-the last terms on the right hand side of  eqs. (28), (29) (for more 
detailed analysis see, e.g Papaloizou $\&$ Lin 1997).
These terms determine the shape of the stationary solutions of the twist equations,
but are not important for the character of time behavior. 
Let us also look for the solution of  eqs. (28), (29) in the WKB-limit:
${\bf W}\propto e^{i(\omega (k) t-kr)}$, $kr < 1$. We obtain the dispersion 
relation $\omega (k)$ in the form (PL):
$$\omega ={1\over 2}
(i\alpha \Omega_{K} \pm \sqrt{k^{2}v_{s}^{2}-\alpha^{2}\Omega_{K}^{2}}),
\eqno 30$$
where $v_s=\delta r\Omega_K$ is the sound velocity. In the case 
$k > {\alpha \Omega_K \over v_{s}}$  eq. (30) describes the propagation of
the waves with velocity $v_W\approx v_s/2$. In the opposite case 
$k < {\alpha \Omega_K \over v_s}$ 
the mode corresponding to the sign $(+)$ in the
eq. (28) describes the decay of the ellipticity of the gas trajectories
due to the action of viscous forces. The mode corresponding to the sign $(-)$
is described by the relation
$$\omega\approx i{\delta^{2}\over 4\alpha}(r^{2}k^{2})\Omega_K. \eqno 31$$
This relation implies that the relaxation to the stationary 
configuration has diffusion-like character, and the characteristic time scale
of this effect $t_{al}(k)\sim \alpha \delta^{-2}{(rk)}^{-2}t_{orb}$. 
Considering the modes with $k\sim r$, it is easy to see that the transition
from wave-like to diffusion-like propagation of the twist corresponds to
$\alpha > \delta$, and when $\alpha > \delta$
the disc alignment time-scale $t_{al}$ can be estimated as:
$$t_{al}\sim t_{al}(k\sim r^{-1})\sim \alpha \delta^{-2}t_{orb}.$$
In the opposite case $\alpha < \delta$, we have:
$$t_{al}\sim r/v_{s} \sim \delta^{-1}t_{orb}.$$ 

In stationary case  eqs. (28, 29) can be written as:
$${d\over dx}{x^{3/2}\over A}{d\over dx}{\bf W}
-{3i\tilde q\over x^{5/2}}{\bf W}=0, \eqno 32$$
$$A=\tilde \alpha -{3i\over 4}\tilde q x^{-2},$$
where we use the explicit form for the precession frequencies (28, 29), 
introduce the dimensionless distance $x={r\over r_{b}}$, and new parameters
$\tilde \alpha={\alpha \over \delta}$ and $\tilde q ={q\over \delta}$.
The eq. (32) is similar to the equation describing a stationary twisted
disc around a Kerr black hole (II) except  for the 
$(-)$ sign  in  front of the forcing term (the last term in (32)). 
Ivanov and Illarionov found  radial oscillations of the inclination angle
 of a low-viscosity stationary twisted disc around a  Kerr black hole, 
and suggested that this effect
holds for any nearly Keplerian forces 
which cause  apsidal and nodal precession of
the disc trajectories provided that these  precessions are in the same
direction.
$({\Omega_{np}\over \Omega_{ap}} > 0)$.
They  also suggested that 
when the precessions counter-rotated $({\Omega_{np}\over \Omega_{ap}} < 0)$
the disc  inclination must join smoothly to the orbital plane of the object.
A twisted disc around a binary system gives a natural example
of  counter-rotating precessions, and we see below in our case that 
the disc lies in the orbital plane for any value of viscosity.

Equation (32) cannot be solved analytically; its numerical solution
is presented at Figs. 2, 3. However, we can obtain  approximate
analytic solutions of
this equation for the case $\tilde \alpha \gg 1$, and  $\tilde \alpha \ll 1$. 
When  $\tilde \alpha \gg 1$ we can set $A\approx \tilde \alpha$, and 
the approximate solution of  eq. (32) 
is expressed in terms of Kelvin 
functions $ker(x)$, $kei(x)$:
$${\bf W}={2^{3/4}{\bf W}_{out}\over \Gamma ({1\over 4})}y_1^{{1\over 4}}
(ker_{1\over 4}(y_{1})+i kei_{{1\over 4}}(y_{1})), \eqno 33$$
where $\Gamma (x)$ is the gamma-function,  
${\bf W}_{out}={\bf W}(r \rightarrow \infty)$, $y_{1}={r_{al1}\over r}$, and
$$r_{al1}={(3\tilde q \tilde \alpha)}^{1/2}r_b. \eqno 34$$
Hereafter we assume ${\bf W}(r \rightarrow 0)\rightarrow 0$ as an inner
boundary condition.
When $r \rightarrow 0$,
$${\bf W}\approx C_{1}(r)e^{-{r_{al1}\over \sqrt {2} r}}
e^{-i({r_{al1}\over \sqrt {2} r}+{3\pi \over 16}-\gamma_{out})}, \eqno 35$$
where $C_{1}(r)$ is a slowly changing real function.
When $\tilde \alpha \ll 1$, $A \approx -{3\over 4}i\tilde qx^{-2}$, and 
the solution is expressed in terms of modified 
Bessel function $K(x)$:
$${\bf W}=2^{3/8}{{\bf W}(out)\over \Gamma (5/8)}y_2^{5/8}K_{5/8}(y_{2}), 
\eqno 36$$
where $y_{2}={({r_{al2}\over r})}^2$, and
$$r_{al2}={{(3\tilde q)}^{1/2}\over 2}r_b . \eqno 37$$
When $r \rightarrow 0$,
$${\bf W}\propto e^{-{({r_{al2}\over r})}^{2}}. \eqno 38$$
Comparing  eqs (35), (38) we see that in the case of a  relatively high
viscosity,  the disc rings are twisting  with  $r$,
but in the  low viscosity limit the effect is absent.
This might give a 
possibility of testing the value of viscosity  in 
circumbinary twisted discs.

The general case $\tilde \alpha \sim 1$ must be analyzed numerically. We
plot the results of numerical 
integration of the twist equation (32)  in Figs.
2,3.
\begin{figure}
\centerline{\epsfxsize=9cm \epsfbox{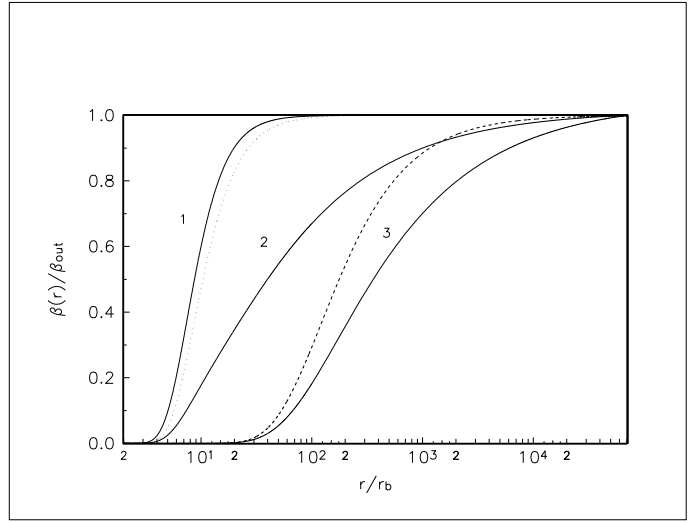}}
\caption{
The dependence of the inclination angle $\beta(r)/\beta_{out}$ on
radius $r/r_b$ for $\tilde q =100$, and for the different values of
$\tilde \alpha$. Curves 1, 2 and 3 correspond to $\tilde \alpha =0.01,
1, 100$ respectively. We also plot two simple analytic approximations
for these curves which correspond to small and large values of $\tilde
\alpha$. The dotted curve $\beta = e^{-{({r_{al2} \over r})}^2}$
($r_{al2}=8.7r_b$), and the dashed curve 
$\beta=e^{-({r_{al1}\over\sqrt{2} r})}$ ($r_{al2}(\tilde \alpha =100)=173.2r_b$) are the
approximate solutions of the twist equation in the limit of high
viscosity (see eqs. 35, 38).}
\end{figure}
\begin{figure}
\centerline{\epsfxsize=9cm \epsfbox{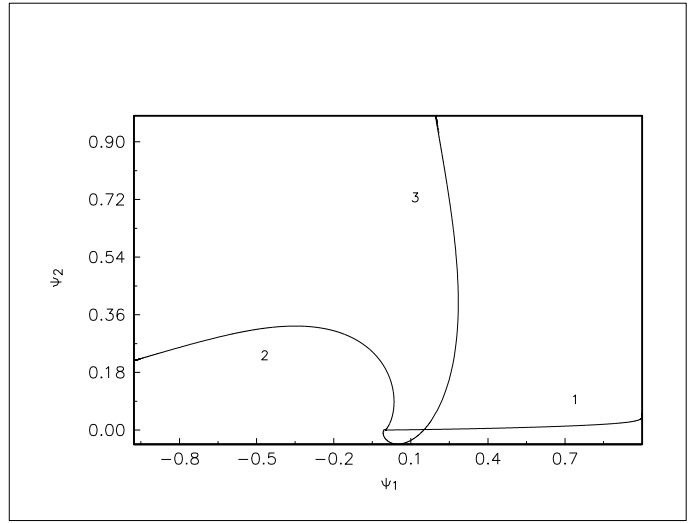}}
\caption{
The result of the integration of the twist equation (32) in a
parametric form ($\Psi_{1}=Re {\bf W}, \Psi_{2}=Im {\bf W}$). The
larger values of $\beta=\sqrt{\Psi_{1}^{2}+\Psi_{2}^{2}}$ correspond
to the larger values of radial distance.  The curve 1 correspond to
the curve 1 of Fig. 2 ($\tilde q=100$ hereafter and $\tilde \alpha
=0.01$), the curves 2 and 3 correspond to the curves 2,3 of Fig. 2
($\tilde \alpha =1$ and $\tilde \alpha =100$ respectively).  When
viscosity is small, the rotational angle $\gamma =\arctan
{({\Psi_1\over \Psi_2})}$ is not changed with the change of $r$, for
larger values of viscosity the rotational angle is changing with the
distance.}
\label{}
\end{figure}

As follows from Fig. 2., the alignment scale decreases with
decrease of the viscosity parameter in agreement with our analytic 
estimates, and the expressions (33), (36) give a good approximation
to the numerical solutions of  eq. (30).
Fig. 3 illustrates the change of the   twist angle $\gamma$ with
the distance $r$. One can see from this Figure  that the angle $\gamma$
  does not change significantly  for small values of the viscosity parameter.

In deriving  eq. (32) we have neglected the self-gravity of the disc.
The effect of self-gravity  is manifest in two different
ways: it gives an  additional non-local forcing term ${\it F}_{sg}$ in 
 eq. (32), as well as a contribution to the resonant denominator $A$.
The term ${\it F}_{sg}$ must be compared with the term induced by
pressure - ${\it F}_{p}$ - the first term on left-hand side of 
eq. (32). According to PL, the forcing term ${\it F}_{sg}$ can be
estimated as: ${\it F}_{sg}\sim {\alpha \over \delta Q}{\it F}_{p}$, 
and we can neglect this term assuming that viscosity takes a  rather
small value
\footnote{In a similar context a self-gravitating twisted disc with  ($Q\sim 1$)
has been recently discussed by  Papaloizou, Terquem and Lin 1998.}
:$$\alpha < \delta Q.$$
On the other hand, the correction to the  resonant denominator is more important
in case of low viscosity, and can be estimated formally by setting
 $\alpha =0.$  As well as  giving a correction to  the Newtonian potential
due to the secondary,  self-gravity induces   apsidal precession of 
 elliptical trajectories with a  frequency $\Omega_{sg}\sim
{M_{d}\over M}\Omega_{K}$. This frequency must be compared to
the frequency $\Omega_{ap}$ at characteristic scale $r_{al2}$:
$$\Omega_{sg}\sim {M_{d}(r_{al2})\over M}\Omega_{K}\sim
{\delta \over Q}\Omega_{K} < \Omega_{ap}(r_{al2})\sim \delta
\Omega_{K}.$$
Thus,  for estimating precession frequencies, the correction due to self-gravity appears to be
less important than the correction due to gravitational field
of the  secondary if $Q > 1$.

At the end of this Section  we note that the low viscosity $\tilde \alpha < 1$
twisted
disc may be unstable with respect to shear instabilities  if the 
``initial'' inclination angle $\beta_{out} $ is sufficiently 
large $\beta_{out} > \delta$ (Coleman $\&$ Kumar 1993, II). 
It was suggested
(II) that the shear
instability might result in development of turbulence 
and  so
effectively increase the value of  the viscosity parameter.
In this case the effective scale of the disc alignment increases
and  equation (37) gives only a lower limit  to the alignment
scale.

\section{Binary  orbit evolution in the  plane of the aligned disc}

Consider now in detail the radial evolution of the disc and the secondary.
In this Section we assume that the disc plane is aligned with
the  orbital
plane of the binary  near the secondary orbit, and the secondary
rotates in the same sense as   the disc.
Hereafter we  adopt the following assumptions: 
a) we assume that the mass of the secondary is  much smaller than the 
primary mass $M$, but larger than the characteristic
disc mass $M_d$
b) the  evolution due to interaction of the secondary and
disc is switched on at some moment of time $t=t_{0}$ when the secondary
lies in the disc at some radius $r_0 \sim r_c < r_d$. 
The disc is taken to have   
a steady state structure  at this moment c) Assuming that the
inner (with respect to the secondary orbit) part of the disc
is relatively quickly swallowed up by the primary, we do not consider
the evolution of that part. 
d)
We assume the coefficient of kinematic viscosity
$\nu$ to be  a power-law function of cylindrical coordinate $r$
and surface density $\Sigma$:
$$\nu=k\Sigma^{a}r^{b}. \eqno 39$$
The expression (39) can be specified for a given model of  the 
accretion disc. In the $\alpha$-disc model the relation
between surface density and viscosity follows from 
consideration of the energy balance. 
Assuming the power-law 
dependence of the opacity $\kappa$ on the density and 
temperature
$$\kappa=\kappa_{0}\rho^{\mu_1}T^{\mu_2},$$
a simple calculation gives (Lyubarskij $\&$ Shakura 1987)
$$k=f{\lbrace \alpha^{8+\mu_1-2\mu_2}\kappa_{0}^{2}
{\cal R}^{2(4-\mu_2)}(GM)^{(2\mu_2+\mu_1-4)/2}
{({\it c} a_r)}^{-2}\rbrace}^{{1\over \epsilon}}, \eqno 40$$
and  
$$a={2(2+\mu_1)\over \epsilon}, \quad b={3(4-\mu_1-2\mu_2)\over 2\epsilon},$$
and $\epsilon=6-2\mu_2+\mu_1$. Here $a_r$ is the radiation density
constant, 
${\cal R}$ is the gas constant,
${\it c}$ is the speed of light, and the standard system of units is assumed.
The parameter $f\sim 1$ depends on the disc structure in vertical
direction and is unimportant for us.  
If  Thompson opacity dominates, $\kappa=\kappa_{Th}=0.4(cm^2g^{-1})$, 
 $a=2/3$, and  $b=1$.  If the opacity
is determined by  free-free processes,
$\kappa_{ff}\approx
0.8\cdot 10^{23}\rho T^{-7/2}(cm^{2}g^{-1})$, and then 
$a=3/7$, $b=15/14$.

An analogous problem was formulated and discussed by Syer and Clarke 1995
(hereafter SC). However, our results differ from those obtained by SC.
As we see below the effect of the interaction between the secondary
and the disc leads to
accumulation of the disc gas  in the  region of the disc
confined between the binary orbital  radius $r_{b}$, and
some 'binary influence' radius $r_{infl}(t) > r_b$. The radius
$r_{infl}$ grows with time. Syer and Clarke approximated $r_{infl}$
as a constant,  and this led to an overestimation
of the disc surface density near the secondary orbit, and
underestimation of the binary evolution time-scale
in their calculations.  
 
\subsection{Evolution of the disc and the 
binary}

The master equation for the
evolution of the disc surface density $\Sigma $ can be written as
(e.g Lightman $\&$ Eardley 1974, Lyubarskii $\&$ Shakura 1987):
$${\partial \over \partial t}\Sigma=
{3k\over r}{\partial \over \partial r}\left(r^{1/2}
{\partial \over \partial r}\left(r^{1/2+b}\Sigma^{1+a}\right)\right). \eqno 41$$
Strictly speaking this equation 
is not valid near the secondary orbit, where the tidal
interaction exchanges the angular momentum between the disc
and the secondary. We assume hereafter that the tidal torque
is concentrated in the narrow disc ring near the secondary
orbit, and  eq.
(41) is valid at all distances outside this ring. Following this
assumption the influence of the secondary on the disc 
can be approximated as 
an effective boundary condition  for eq. (41),
which takes into account
the exchange of the $z$-component of the angular 
momentum between the secondary and the disc 
(SC)
$$\lbrace 2\pi r^{2}\Sigma({3\over 2}\nu \Omega_K+ v_{dr}r\Omega_K)+
{1\over 2}mr\Omega_K {dr_b\over dt}\rbrace_{r=r_{b}}=0, \eqno 42$$
where all quantities are taken at $r=r_{b}(t)$, and
the secondary radial velocity ${dr_b\over dt}$
is of order of the disc gas drift velocity $v_{dr}$ 
near the secondary orbit. In general, 
$$v_{dr}=-{3\over \Sigma r^{1/2}}{\partial \over \partial r}\left(
r^{1/2}\nu \Sigma \right), \eqno 43$$
In the case of 
a massive secondary ($m \gg M_d$),
the second term in the round brackets 
can be neglected.

Prior to the moment $t=t_{0}$, we set ${\partial \Sigma \over
\partial t}=0$ into  eq. (41). In this steady state case 
we have
$$\eta_{0}\equiv \nu_{0}\Sigma_{0}={\dot M\over 3\pi}, \eqno 44$$
$$\Sigma_{0}={(3k)}^{-{1\over 1+a}}
{({\dot M\over \pi})}^{{1\over 1+a}}r^{-{b\over 1+a}}, \eqno 45$$
$$v_{r0}=-{1\over 2}{(3k)}^{1\over 1+a}{({\dot M\over \pi})}^{{a\over 1+a}}
r^{{b-a-1\over 1+a}}, \eqno 46$$
$$t_{\nu 0}={r\over |v_{r0}|}
=2{(3k)}^{-{1\over 1+a}}{({\dot M\over \pi})}^{-{a\over 1+a}}r^{{c
\over 1+a}}, \eqno 47$$
where 
$$c=2(a+1)-b.$$
We assume below that $c > 0$.

The presence of the secondary in the disc significantly influences 
the structure of the inner part of the disc. Here we describe this
influence in the most crude approximation ( a more
accurate approach is discussed in the next subsection). 
The main effect of the secondary is to
slow down the radial motion in the disc: $v_{r} \ll v_{r0}$. 
To describe this effect quantitatively we divide the disc by
two regions: 1) the outer region $r > r_{inf} \gg r_{b}$, 
where the equations
(44-47) are approximately valid,  
2) the inner region $r_{b}< r < r_{inf}$,
where the drift velocity (43) is  small compared to 
the stationary value (46). 
These regions are separated by the 'influence'
radius $r_{infl}(t)$, and the dependence 
of $r_{infl}$ on time 
is calculated below. When  $v_{r}\approx 0$,
 eq. (43) gives 
$$\Sigma(r, t)=C(t)r^{-{b+1/2\over a+1}}. \eqno 48$$
The functions $C(t)$, and $r_{infl}(t)$ are calculated using
the laws of conservation of mass and angular momentum.
The eq. of mass conservation has a form
$$\int_{0}^{r_{infl(t)}}dr({\partial \over \partial t} \Sigma r)=
{\dot M\over 2\pi}, \eqno 49$$
and the eq. of conservation of angular momentum has a form
$$\int_{0}^{r_{infl(t)}}dr({\partial \over \partial t}\Sigma
r^{3}\Omega)=
\lbrace {3\over 2}r^{2}\Omega\nu\Sigma \rbrace_{r\rightarrow 0}.
\eqno 50$$
In both equations we formally extend 
the lower limit of integration to
$r=0.$ In  eq. (50)  we neglect the small inflow of angular 
momentum from the outer (steady-state) region of the disc. 
After substitution of
(47) into eqs. (49-50), we have
$$\dot Cr_{infl}^{{2c-1\over 2(a+1)}}=\lbrace {2c-1\over a+1}\rbrace
{\dot M\over
4\pi}, \eqno 51$$
$$\dot Cr_{infl}^{{2c+a\over 2(a+1)}}={3\over 4}\lbrace {2c+a\over a+1}
\rbrace kC^{a+1}.
\eqno 52$$
The solution of eqs. (51-52) up to unimportant constant of integration
has  the form
$$C={(3k)}^{-{2c-1\over 2(a+1)c}}
{({\dot M \over \pi})}^{{2c+a\over 2(a+1)c}}
{(\beta t)}^{{1\over 2c}} \eqno 53$$ 
$$r_{infl}={(3k)}^{1\over c}
{({\dot M\over \pi})}^{{a\over c}}{({2c+a\over 2c-1})}^{2}
{(\beta t)}^{{a+1\over c}}, \eqno 54$$
where
$$\beta={({c(2c+a)\over 2(a+1)})}{({2c+a\over
 2c-1})}^{-{2c+a\over (a+1)}}.$$
Prior to the moment $t_0$ the surface density in the disc is
given by the steady-state solution (45), and after that moment, 
the surface density should be described by the solution (48, 53). 
To reconcile these two solutions  we set
$$t_0={ t_{\nu 0}(r_0)\over 2\beta},$$
where $t_{\nu 0}(r_0)=t_{\nu 0}(r=r_0)$. 
In this case  eqs. (48), (53) can be written
as
$$\Sigma=\Sigma_{0}(r_{0}){(t/t_{0})}^{{1\over 2c}}
{(r/r_{0})}^{-{b+1/2\over a+1}}, \eqno 55$$
and for the viscosity distribution we have
$$\eta={\dot M\over 3\pi}{(t/t_{0})}^{{a+1\over 2c
}}{(r/r_{0})}^{-1/2}, \eqno 56$$
where $\eta=\nu \Sigma $.
The expressions (55-56) are valid  as long as the local viscous time $t_{\nu}
={r^{2}\over \nu}$ doesn't exceed 
the characteristic dynamical time
$t_d={\Sigma \over \dot \Sigma} \sim t$. We have
$$t_{\nu}/t\sim {(r/r_{infl})}^{{2c+a\over 2(a+1)}}, \eqno 57$$
so our approximation  breaks down  at $r\sim r_{infl}$.
Thus, to describe the disc structure at $r\sim r_{infl}$ we 
need  a more accurate approach
(see next subsection).

Now we are able to calculate the evolution of the binary orbit.
First, let us calculate the secondary radial velocity 
${dr_{b}\over dt}$.
For that we use  eq. (42), and then neglecting the second term in the
brackets we obtain
\footnote{In the  case of a  relatively massive secondary,  the inner edge
of the disc is determined by a few outer Lindbland resonances,
and may be 2-3 times larger than the secondary
orbital  radius (e.g. Artymowicz $\&$ Lubow 1994). However we can
use  eq. (58) to calculate the  radial velocity. This is connected
with fact that the angular momentum outflow does not depend on
the distance at all, see eq. (50).} 
$${dr_{b}\over dt}=-{6\pi r_{b}\eta(r_{b}, t)\over m}, \eqno 58$$
and with help of  eq. (56)  eq. (57) can be rewritten 
in dimensionless form 
$${dy_b\over d\tau}=-{S\over 2\beta}\tau^{{a+1\over 2c}}, \eqno 59$$
where $\tau=t/t_{0}$, $y_b={(r_b/r_{0})}^{1/2},$ 
and dimensionless parameter
$$S={M_{d0}\over m} < 1,$$
here $M_{do}=\dot M t_{\nu 0}(r_0)$. 
The integration of
 eq. (59) gives
$$r_b={(1- \gamma S (\tau^{{5c+b\over 4c}}-1))}^{2}r_{0}, 
\eqno 60$$
where $\gamma= {2c \over \beta (5c+b)}\sim 1$.
The characteristic time-scale of the evolution of the secondary
orbit $t_{ev}$ can be found from the condition $r_b(t_{ev})=0$:
$$t_{ev}= 
{(\gamma S )}^{-{4c\over 5c+b}}t_{0}. \eqno 61$$
It is instructive to compare the expression (60) with  the 
Syer and Clarke result: 
$$r_{b}\approx {(1-\beta_{SC}S^{{1\over 2+a}}\tau)}^{{2+a\over c}}r_{0},
\eqno 62$$
where the dimensionless coefficient 
$\beta_{SC}=2^{-{1+a\over 2+a}}{c \over \beta (2+a)} \sim 1$. 
It can be seen that for the typical values of the 
parameters $a, b$ the evolution equation (62) underestimates
the evolution time in comparison with  equation (60).

The dimensionless velocity (59) taken at the time $t_{ev}$
can be estimated as:
$$|{dy_b\over d\tau}| \sim S^{4c\over 5c+b} < 1. $$
This inequality implies that we can treat the change of binary orbit
as a slow effect with respect to the change of the disc configuration
when $t < t_{ev}$. 
On the other hand, let us look at the ratio of the 
disc mass contained inside the region 
influenced by the
secondary ($r_{b} < r < r_{infl}$) to the mass of the secondary.
It can be easily seen that this ratio
is small at the time $t\sim t_{ev}$:
$$M_{d}=
2\pi \int_{r_{b}(t_{ev})}^{r_{infl}(t_{ev})} drr \Sigma (t_{ev}, r)
\sim  S^{{2(a+1)\over 5c+b}}m <  m,
\eqno 63$$
and this means 
we can use  eq. (58) as a boundary condition to our problem.
It is useful to express  the evolutionary time $t_{ev}$ in 
terms of the accretion time $t_{acc}={m\over \dot M}$. From  eq. (61)
we obtain
$$t_{ev} ={(\gamma S)}^{{2(a+1)\over 5c+b}}t_{acc} < t_{acc}. \eqno 64$$
Say, for $a=2/3$, $b=1$, we have
$$t_{ev}={(\gamma  S)}^{5/19}t_{acc}, \eqno 65$$
and for $a=3/7$, $b=15/14$, we have
$$t_{ev}={(\gamma S)}^{2/7}t_{acc}. \eqno 66$$

Finally let us estimate the luminosity of the disc.
From the law of energy conservation
it follows that the total luminosity of the disc $L$
must be equal to the change of the secondary energy per unit of time:
$$L=-{1\over 2}{GmM\over r_{b}^{2}}{dr_{b}\over dt}
={GM\dot M\over r_{b}(t)}{({r_{0}\over r_{b}(t)})}^{1/2}
{(\tau )}^{{a+1\over 2c}}, \eqno 69$$
where $r_{b}(t)$ is given by  eq. (60).   
When the separation distance $r_{b}$ 
is much larger than the
gravitational radius of the primary, 
the ``initial'' 
value of luminosity (taken at the moment $t=t_{0}$) is much smaller than
the ``standard'' value of luminosity of the steady state disc $L_{st}$
$$L_{st}\sim {\dot M M\over r_{g}}\approx
10^{44}\dot M_{*}M_{8}{erg\over s}. $$ 
However, the luminosity grows with time
due to evolution of inner part of the disc and the secondary orbit.

\subsection{Self-similar solutions}

Now we consider in detail 
the formal case 
of infinitely small binary separation distance $(r_{b}\approx 0)$.
As we will see, in this case  simple self-similar solutions
of  eq. (41) can be found
\footnote{Our approach is similar to the  Lyubarskii and Shakura 1987
study of non-stationary disc accretion, and to Pringle's 1991
calculation of
evolution of external circumbinary disc of a given mass.}. 
These solutions are close to the
exact solutions for the disc structure outside the binary orbit,
and they allow us to find the disc structure at the intermediate 
scales $r\sim r_{infl}$.

When considering the case $r_{b}=0$, we must specify the boundary conditions
at $r=0$ and at infinity. We assume that there is no radial velocity
at $r=0$: $v_{r}(r=0)=0$, and the disc structure tends to the steady
solution with increase of $r$. 
We look for a solution of  eq. (41) in the form
$$\sigma=t^{-m}F(\xi), \eqno 70$$ 
$$\xi=D^{-1}rt^{-n}, \eqno 71$$
where we use the surface number density $\sigma=\Sigma/m_{p}$
($m_{p}$ is the proton mass),
and the constant $D$ is chosen to make the similarity  variable (71)
dimensionless. The powers $n$, $m$
are specified by the condition that  eq. (41), as well as 
the accretion rate
$${\dot m\over 2\pi}\equiv {\dot M\over 2\pi m_{p}}=
3r^{1/2}{\partial \over \partial r}\left(r^{1/2}\nu \sigma\right), \eqno 72$$
are expressed in terms of the variable $\xi$ only. Substituting
 eqs. (70) and (71)  into eqs. (41) and (72), we obtain
$$n={1+a\over c}, \quad m={b\over c}. \eqno 73$$
Equation (41) takes the form
$$(3\tilde k)D^{b-2}{d\over d\xi}\xi^{1/2}{d\over d\xi}\xi^{1/2+b}F^{1+a}
+\xi(n\xi{d\over d\xi}F+mF)=0, \eqno 74$$
with 
$$D=(3\tilde k)^{{1\over c}}{({\dot
m \over \pi})}^{{a\over c}}, \eqno 75$$
where $\tilde k =m_p^a k$.
The solution of eq. (74) should be separately analyzed for the
general situation $a\ne 0$, and for the degenerate
case $a=0$. 

In the case $a=0$ we have $(3\tilde k)D^{b-2}=1$, Eq. (74) is linear,
and can be easily solved. The general solution is
$$F=\xi^{-b}(C_{1}+C_{2}\int^{\xi^{-1/2}}_{0}dx e^{-{x^{-2(2-b)}\over
{(2-b)}^{2}}}). \eqno 76$$
The first term in the brackets represents the stationary solution
(45) provided $C_{1}=1$. We have
$$F\approx C_{2}\xi^{-b-1/2}, \eqno 77$$
at $\xi \rightarrow 0$, and the
comparison of  eqs. (70,71), (77),
and (48), (53) gives $C_{2}=\beta $.

In the case $a\ne 0$, we can eliminate dimensional factor in the 
eq. (74) by redefining of the function $F$
$$F=(3\tilde k)^{-{1\over a}}D^{{2-b\over a}}\tilde F. \eqno 78$$
Substituting (73) (77) into (74), we obtain:
$${d\over d\xi} \xi^{1/2} {d\over d\xi}\xi^{1/2+b}{\tilde F}^{1+a}
+{\xi \over c}(b\tilde F+(1+a)\xi{d\over d\xi}\tilde F)=0.
\eqno 79$$
In the limit $\xi \gg 1$
the solution of eq. (79) is close to the stationary solution
$$\tilde F_{0}=\xi^{-{b\over 1+a}}, \eqno 80$$ 
and in the opposite limit $\xi \ll 1$ we have
$$\tilde F=C\xi^{-{b+1/2\over 1+a}}.  \eqno 81$$
It is evident from  eqs. (70, 71, 81)  that
the asymptotic (81) corresponds to the approximate
solution (48, 53) if we set again $C=\beta$.
Thus  eqs. (45) and (48, 53)  give asymptotic  limits 
to the self-similar solution of  eq. (41).  
The intermediate scale $\xi \sim 1$
separates the asymptotic (80,81), and
the corresponding radius $r_{*}(\xi=1, t)$ is of order of $r_{infl}$.

The numerical solution to  eq. (79) is presented in Fig. 4.
\begin{figure}
\psfig{figure=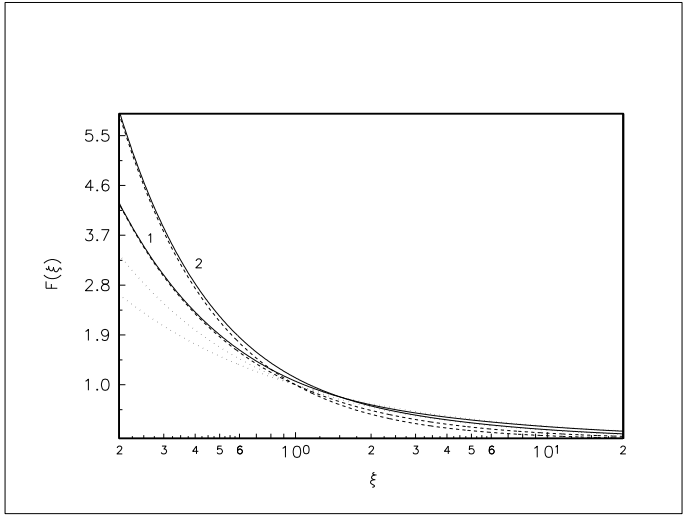,width=9cm}
\caption{We plot the numerical solution of  eq. (79) $F(\xi)$ as
a function of the similarity  variable $\xi$ (solid curves). 
 Curve 1 corresponds
to the case $a=2/3$, $b=1$ (Thompson opacity dominates), and the 
curve 2 corresponds to the case $a=3/7$, $b=15/14$ (the
opacity is determined by the free-free processes). The dashed lines
represent the analytical asymptotic  of that solutions at $\xi \rightarrow
0$ (see eq. (81)), 
and the dotted lines represent the stationary solutions of  eq. (79)
(see eq. (80)) }
\label{}
\end{figure}
One can see from this Fig., that the solution of this equation
approximated with high accuracy by the expression (81) at
$\xi < 1$, and by the expression (80) at $\xi > 1$.
We will use this fact below comparing our analytical estimates
with the results of the numerical computations.

\subsection{Numerical model}

We solve numerically  equation (41), and compare the results of 
our calculations with analytical estimates. For numerical work we rewrite 
 eq. (41) in terms of new radial coordinate $y={(r/r_0)}^{1/2}$, 
dimensionless time $\tau=t/t_0$, and dimensionless surface density
$\tilde \Sigma =\Sigma /\Sigma_{0}(r_{0})$:
$${\partial \tilde \Sigma \over \partial \tau}=
{1\over 4\beta}y^{-3}{\partial^{2}\over \partial y^{2}}
(y^{1+2b}{\tilde \Sigma}^{1+a}). \eqno 82$$
To solve this equation,
we must specify the initial and boundary conditions. We assume
that the disc surface density is equal to its steady state value  
(45) at some point $y_{out}\gg 1$, and we take
$y_{out}=10^{2}$ in our calculations. Since  we are 
interested in the effect of evolution of the disc
and the secondary orbit, the
detailed description of the angular momentum exchange between
the secondary and the disc is unimportant for us, 
and to obtain a boundary condition at $y=y_{b}$
we substitute the expression for the disc gas radial
velocity (43) in  eq. (42) and use the resulting equation
as the boundary condition to our problem.
The eq. (43) allows us to calculate the change of
the secondary orbit. The initial surface density distribution
$\Sigma (t=0, y)$ is determined by eq. (45) in the range
$2 < y < y_{out}$, and we use the appropriate polynomial 
function in the range $1 < y < 2$ to match the steady state
distribution and the 
inner boundary condition at the moment $t=0$. The detailed form
of this function is unimportant for us.
The position of the secondary $y_{b}={(r_{b}/r_0)}^{1/2}$ is changing with
time, and in our numerical scheme it is convenient to use another 
radial coordinate $\tilde y$ which is stretching with change 
of the inner boundary with respect to coordinate $r$:
$$\tilde y={(y_{out}-1)y+y_{out}(1-y_b(t))\over 
y_{out}-y_{b}(t)}, \eqno 83$$
and $\tilde y(y_{b}(t))=1$, $\tilde y(y_{out})=y_{out}$.
We use uniform mesh in coordinates $(\tau, \tilde y)$ with $10^{4}$
grid elements in $\tilde y$-direction, and employ a standard
implicit scheme.
To check the relaxation of numerical solution to the self-similar
asymptote, we integrate  eq. (82)  with respect to time up to the time
$t_{out}=2 t_{ev}$. When the orbital distance approaches the
value $y_{in}=10^{-2}$ we ``turn off'' the radial evolution of
the secondary.

The results of numerical calculations are presented at Figs. (5-7).
In Figs. 5,6 we show the calculated density profile taken 
in different moments of time, and for different parameters of our
model. One can see from these Figs. that the surface density of the
disc is well approximated by the analytical expressions (45), (55).
In Fig. 7 we show the dependence of the secondary orbital distance
on time. The analytical curves give again very good approximation
to the numerical results.
\begin{figure*}
\parbox{16cm}{\centerline{
\psfig{figure=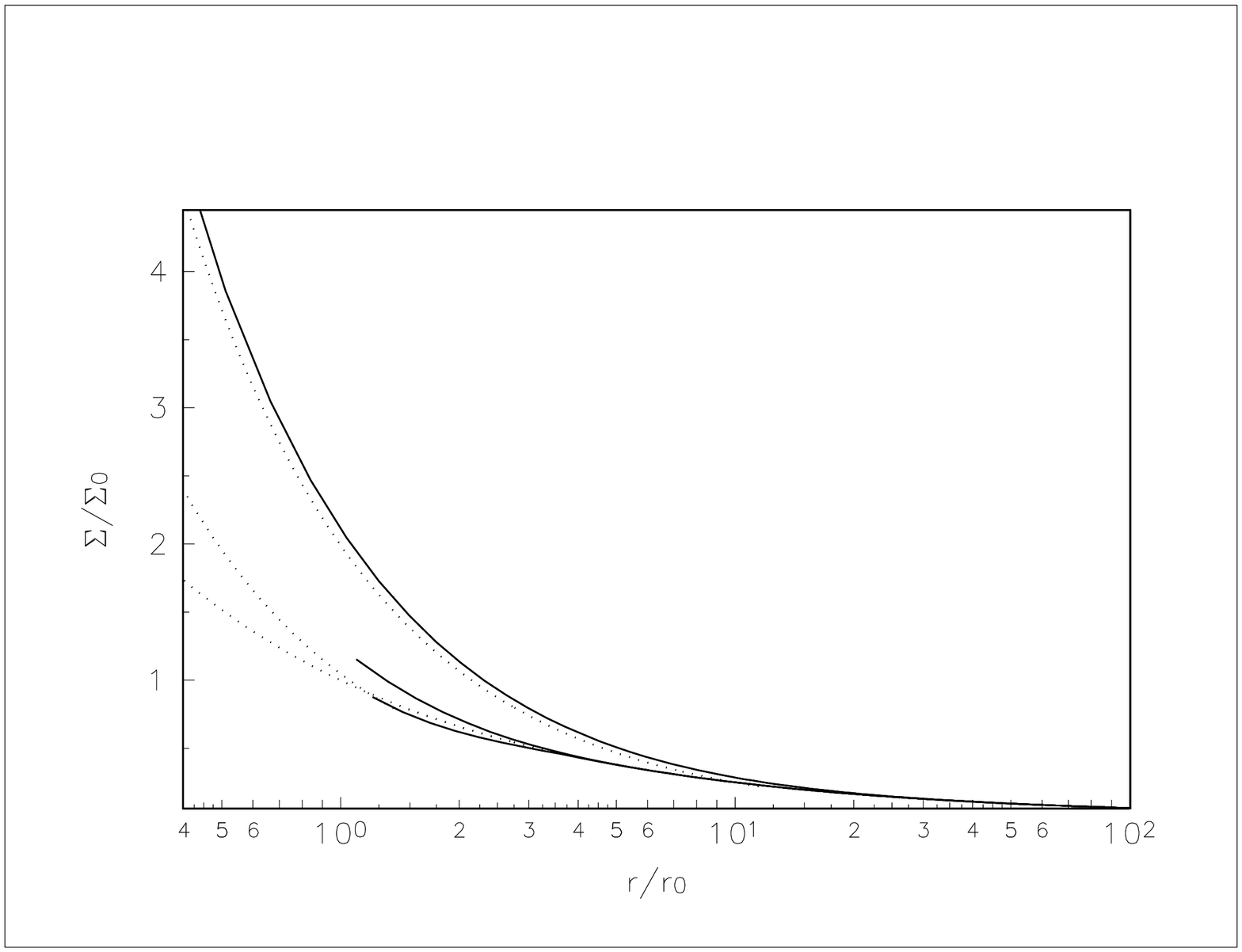,width=9cm,angle=-90}  
\psfig{figure=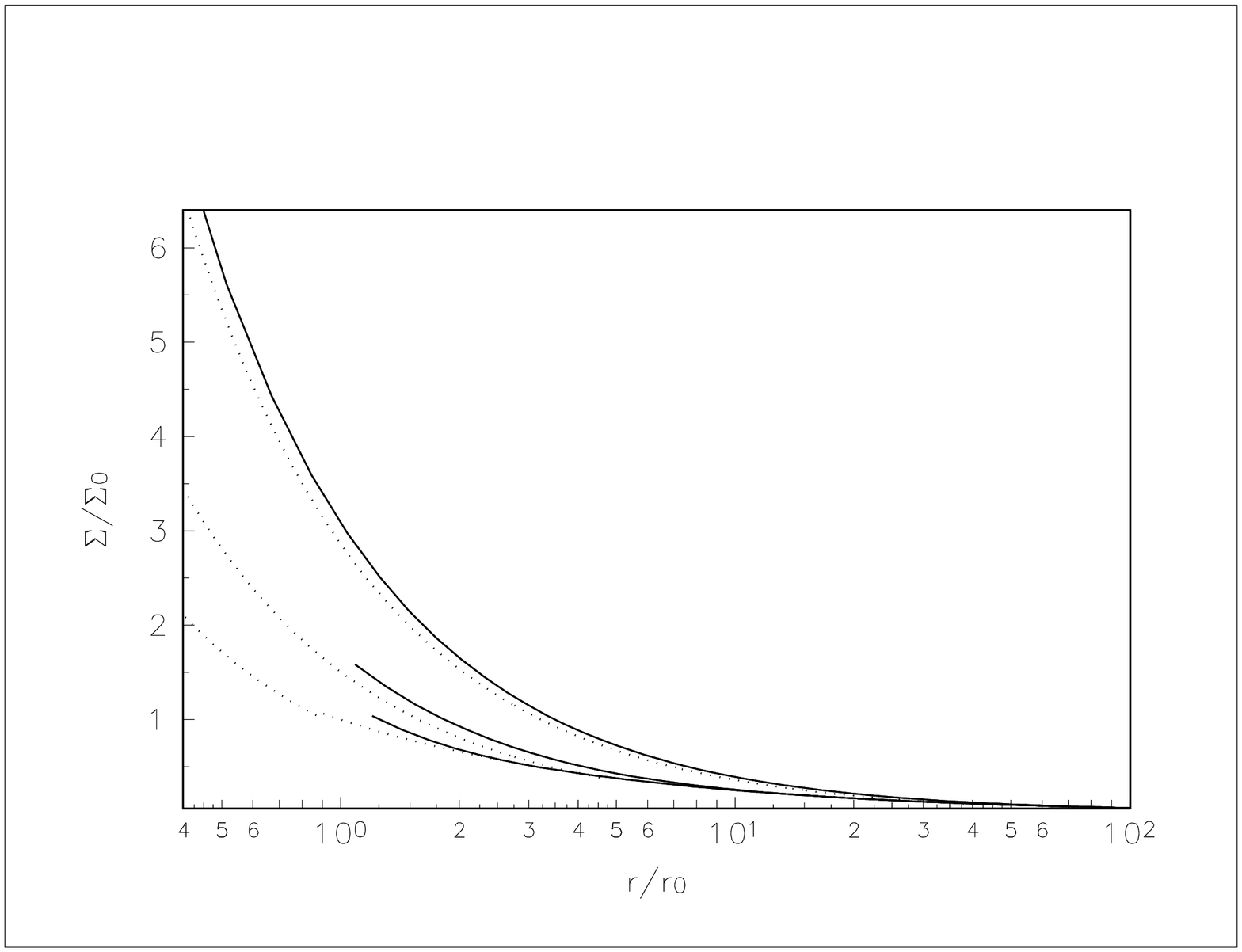,width=9cm,angle=-90}
}}
\parbox{16cm}{\centerline{
\psfig{figure=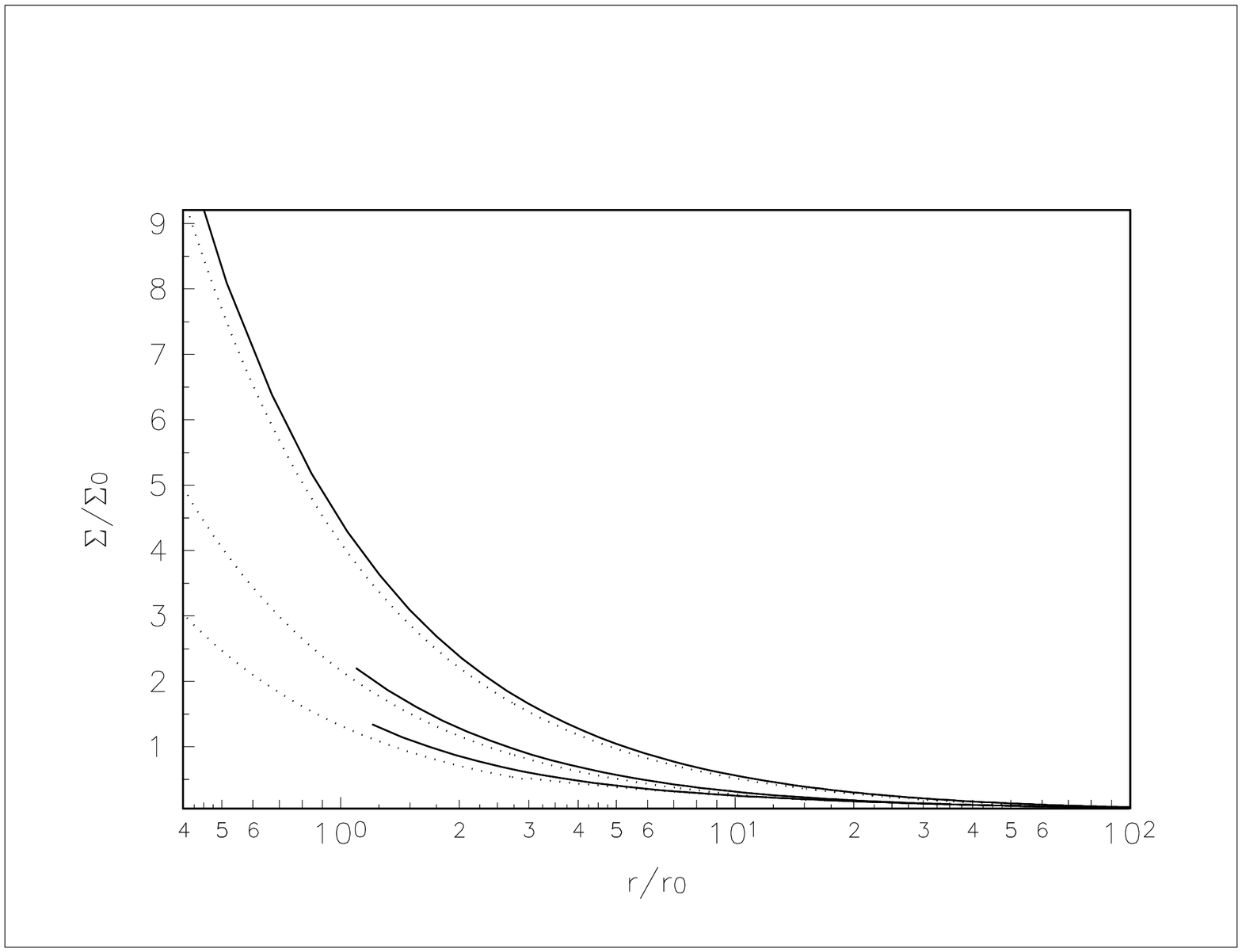,width=9cm,angle=-90}
\psfig{figure=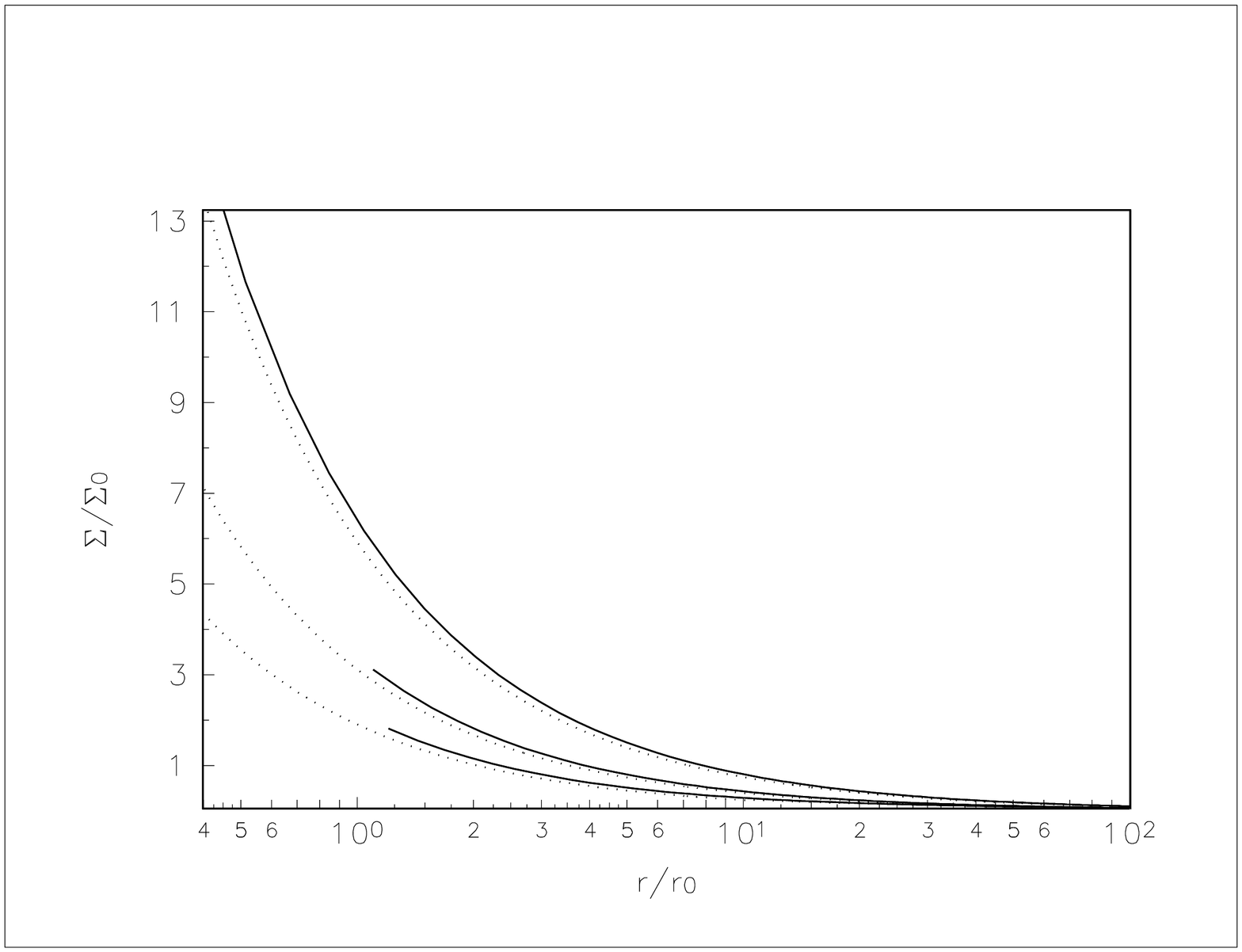,width=9cm,angle=-90}
}}
\caption{The dimensionless surface density $\Sigma/\Sigma_{0}(r_{0})$
is plotted against dimensionless radius $r/r_{0}$. The parameters
$a=2/3$ and $b=1$ correspond to the case of Thompson opacity ``domination''.
The different plots correspond to the different values of the parameter
$S$: a) top left plot -$S=10^{-1}$, b) top right plot-$S=10^{-2}$, 
c) bottom left plot -$S=10^{-3}$, d) bottom right plot-$S=10^{-4}$.
The solid curves, in order of increasing of inner boundary, are made
at times $t=10^{-2}t_{ev}$, $t=10^{-1}t_{ev}$ and $t=2t_{ev}$. The dotted
curves are determined by eqs. (45,55).
In asymptotics the solutions have power-law form:
$\Sigma (r\rightarrow \infty )\sim r^{-3/5}$, and 
$\Sigma (r\rightarrow 0 )\sim r^{-9/10}$. }
\end{figure*}
\begin{figure*}
\parbox{16cm}{\centerline{
\psfig{figure=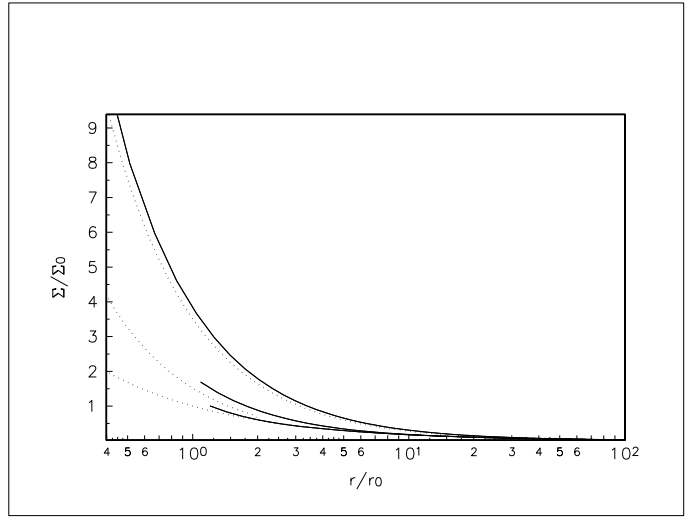,width=9cm,angle=-90}}}
\caption{The same as the plot b) of Fig. 5, but now $a=3/7$, $b=15/14$ 
(the opacity is determined by free-free processes). 
The general behavior of the curves is similar to the case b), but
the asymptotical slope is slightly 
steeper: 
$\Sigma (r\rightarrow \infty )\sim r^{-3/4}$, and 
$\Sigma (r\rightarrow 0 )\sim r^{-11/10}$}
\end{figure*}
\begin{figure*}
\parbox{16cm}{\centerline{
\psfig{figure=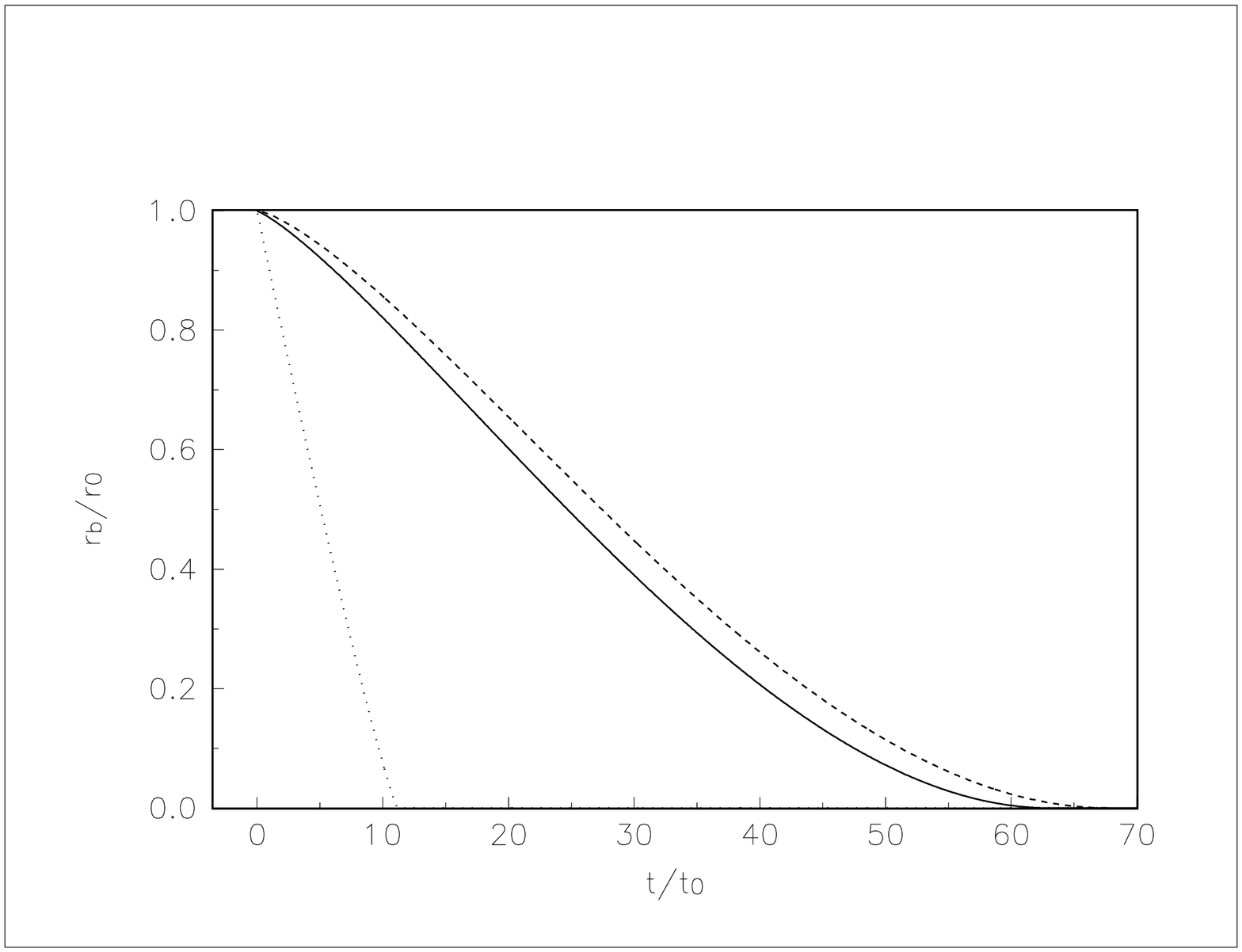,width=8cm,angle=-90}
\psfig{figure=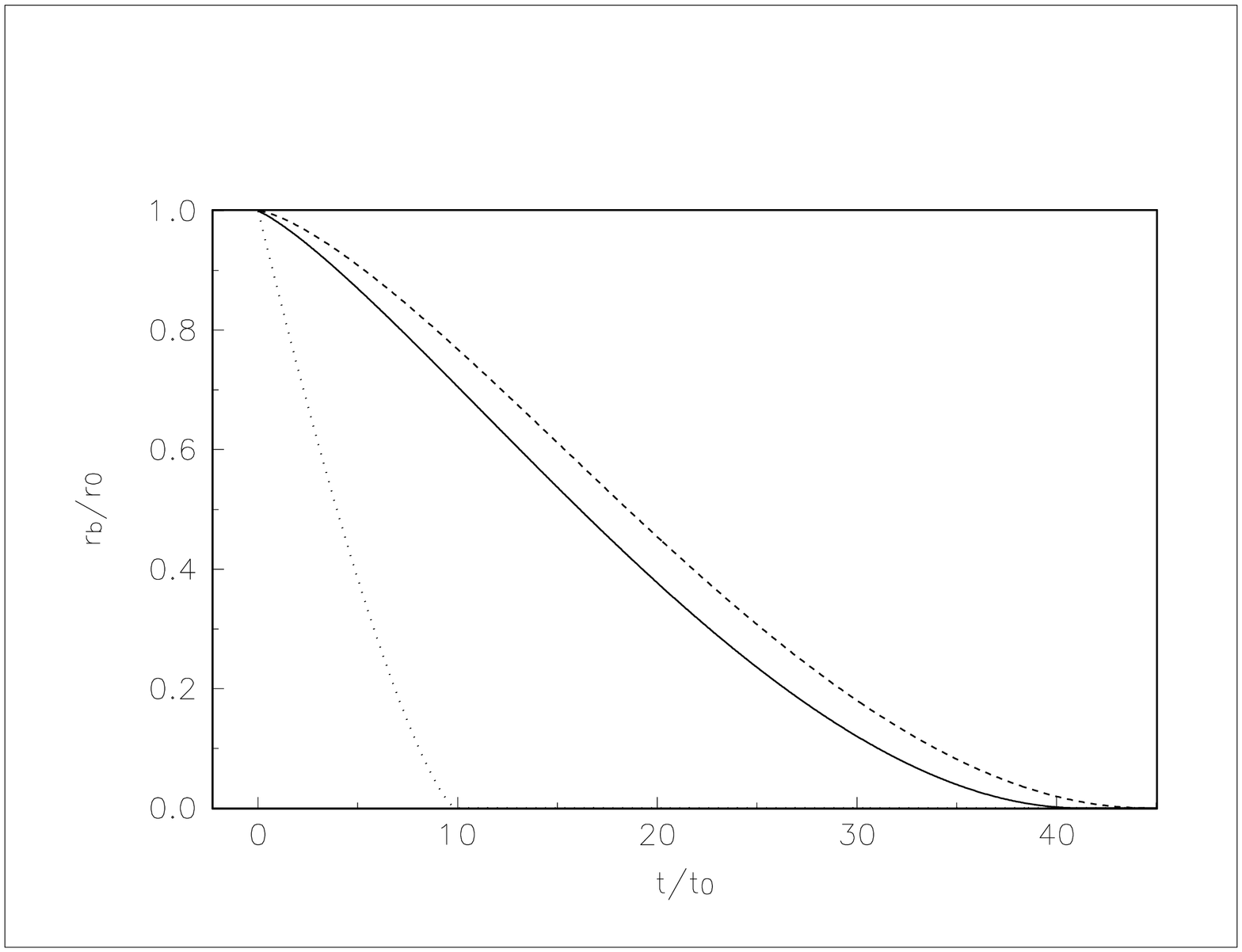,width=8cm,angle=-90}
}}
\caption{The dimensionless binary separation distance $r_{b}/r_{0}$ is
plotted against the time $t/t_{0}$. The parameters of the left plot are
the same as in the plot b) of Fig. 5, and the parameters of the 
right plot are the same as in Fig. 6. 
The dashed curves are analytical expressions
given by eq. (60), and the dotted curves represent Syer and Clarke 
result (62) }
\end{figure*}

\section{Astrophysical consequences and conclusions}

Now let us summarize the main results of our paper. We considered 
the interaction between a massive binary system and an accretion
disc,  paying special attention to the case when the  secondary mass
is larger than the disc mass inside the binary  orbit.
 We showed that the joint evolution of the binary and
disc could  be divided  into several successive stages. 
When the binary orbital plane is inclined with respect
to plane of the disc, the gravitational field of the binary
causes non-axisymmetric ``twisted'' distortions of an initially
planar disc, and the inner part of the disc lies in the orbital 
plane as a result of  the evolution of these distortions. Since the
orientation of outer part of the disc is controlled by the
disc formation processes, after the alignment of the inner part
of the disc  a quasi-stationary twisted disc is formed.
We calculate the radial dependence of the inclination angle
of the twisted disc, and find the dependence of the characteristic
alignment scale on viscosity and  mass ratio of the binary
components. It is shown that in the low-viscosity limit  
the alignment scale does not depend
on the value of  the viscosity parameter $\alpha$.

Assuming that the inner part of the disc lies in the binary orbital
plane, rotates  in the same sense as the binary,  we find the radial evolution
of the disc structure and the evolution 
of the binary separation distance. In such a  situation the disc gas
cannot drift through the binary orbit, and is accumulated in some
region outside the binary orbit. The structure of the disc in this
region is close to the self-similar solution of the  diffusion equation
describing  the viscous evolution of the disc surface density. 
The back reaction of the circumbinary disc pushes the secondary
inwards towards the primary. We calculate analytically the dependence of
the binary separation distance on time, and compare with results
of numerical calculations. We find that the characteristic time
scale of this evolution is determined by  the ratio $S$ of the
disc mass inside the initial binary orbit and the secondary
mass $m$. The evolution time scale is always smaller than
the ``accretion'' time scale $t_{acc}={M\over \dot M}$. 

In  the case of  a supermassive black hole binary,
 evolution of the binary separation distance is also
caused by the emission of gravitational waves. To estimate the relative
significance of this effect, let us compare the total
disc luminosity
with orbital energy loss $\dot E_{gw}$ due to emission of
gravitational waves (Peters $\&$ Mathews 1963)
\footnote{In the expression (84) we set the orbital eccentricity
$e=0$. For $e\sim 1$ the emission of gravitational waves is much more
effective (Peters $\&$ Mathews 1963).}: 
$$\dot E_{gw}={1\over 5}L_{*}q^{2}{({r_g\over r_{b}})}^{5},
\eqno 84$$
where the  ``gravitational emission factor'' $L_{*}={c^{5}\over G}\approx
3.6\times 10^{59}erg/s$. Comparing  eqs. (69), and  (84), we find that
the interaction with the  accretion disc is more important at scales
larger than  the  characteristic ``gravitational'' scale $r_{gw}$:
$$r > r_{gw}\approx 6\cdot 10^{2} {({q\over 10^{-2}})}^{1/2}
{(M_{8}\dot M_{*})}^{-1/4}r_{g}. \eqno 85$$
As we have already discussed, the evolution of  a supermassive black 
hole binary due to dynamical friction with the  stars of the central
stellar cluster can be unimportant at scales $< 1pc$, and 
therefore the secondary-disc interaction may provide the only 
possibility to pass through  the ``intermediate'' scales $r_{gw} <
r < 1pc$. This fact may be critically important for formation
of powerful sources of gravitational radiation due to coalescence 
of supermassive black holes.

 One  uncertainty in our calculation seems to be connected with
eccentricity evolution. In the  course of binary evolution due to
dynamical friction with the   stars, the eccentricity seems not to grow
significantly (see, e. g. Polnarev $\&$ Rees 1994 for analytic  work and 
Quinlan $\&$ Hernquist 1997 for N-body simulations). However,
the binary-disc interaction may induce 
some eccentricity (e.g. Artymowicz, 1992, Lin \& Papaloizou, 1993). As we have
mentioned,  gravitational wave emission is much more effective
for eccentric binaries, and  a significant  eccentricity
can shorten the evolutionary time scale.

\section*{{\bf ACKNOWLEDGMENTS}}
We thank N. I. Shakura and C. Terquem for useful comments. We are also grateful to
the referee for important remarks. PBI and JCBP
thank  the Isaac Newton Institute for Mathematical Sciences for hospitality.
This work was supported in part 
by RFBR grants N 96-02-16689a, 97-02-16975 
and by Danish Research Foundation
through its establishment of the Theoretical Astrophysics Center.

\bsp

\label{lastpage}

\end{document}